%% file: main.tex
\documentclass[sigplan,10pt,nonacm]{acmart}
\renewcommand\footnotetextcopyrightpermission[1]{}
\usepackage{booktabs}   

\usepackage{algorithm}
\usepackage{algorithmicx}
\usepackage[noend]{algpseudocode}

\usepackage{graphicx}
\usepackage{subfigure}
\usepackage{multirow}
\usepackage{enumitem}

\input{command.tex} 

\begin{document}
\pagestyle{plain}

\title[]{\name: Enabling Learning On-device Contiguously for Agent LLMs} 


\author{Xinxin Liu}
\affiliation{
    \institution{Nanjing University}
    \city{Nanjing}
    \country{China}
}
\email{xinxinliunju@gmail.com}

\author{Jiaxin Li}
\affiliation{
    \institution{Nanjing University}
    \city{Nanjing}
    \country{China}
}
\email{jiaxin66778@gmail.com}

\author{Zibo Wang}
\affiliation{
    \institution{Nanjing University}
    \city{Nanjing}
    \country{China}
}
\email{wangzb@smail.nju.edu.cn}

\author{Yun Ji}
\affiliation{
    \institution{Nanjing University}
    \city{Nanjing}
    \country{China}
}
\email{jiyun@smail.nju.edu.cn}

\author{Zhangqi Zhu}
\affiliation{
    \institution{Nanjing University}
    \city{Nanjing}
    \country{China}
}
\email{231880166@smail.nju.edu.cn}

\author{Qing Hu}
\affiliation{
    \institution{Nanjing University}
    \city{Nanjing}
    \country{China}
}
\email{231880407@smail.nju.edu.cn}

\author{Zhibin Wang}
\authornote{Corresponding author.}
\affiliation{
    \institution{Nanjing University}
    \city{Nanjing}
    \country{China}
}
\email{wzbwangzhibin@gmail.com}

\author{Rong Gu}
\affiliation{
    \institution{Nanjing University}
    \city{Nanjing}
    \country{China}
}
\email{gurong@nju.edu.cn}

\author{Sheng Zhong}
\affiliation{
    \institution{Nanjing University}
    \city{Nanjing}
    \country{China}
}
\email{sheng.zhong@gmail.com}

\author{Chen Tian}
\affiliation{
    \institution{Nanjing University}
    \city{Nanjing}
    \country{China}
}
\email{tianchen@nju.edu.cn}

\renewcommand{\shortauthors}{Liu et al.}

\begin{abstract}

On-device LLM agents interact repeatedly with users on local hardware,
producing private traces that are valuable for adaptation but should not be
sent to a remote trainer. Ideally, such agents would learn
\emph{contiguously}---adapting from every interaction without pausing or
suspending user-facing inference---yet existing inference runtimes assume
stable weights and existing RL systems assume separated resources, so neither
can support this continuity. We present \name, the first single-GPU runtime
that enables contiguous on-device learning for LLM agents. The key insight is
that GPU scheduling, adapter version management, and KV-cache validity cannot
be handled by independent subsystems: adapter updates invalidate cached KV
tensors from older versions, and cache retention affects the memory available
for training. \name makes adapter version, task priority, and cache state
visible to three cooperating components---a cooperative scheduler, a
version-aware KV-cache manager, and a multi-agent model runtime---that share
this state to keep scheduling, execution, and cache maintenance mutually
consistent. On a single 24\,GB GPU with 7B-class models, \name lowers
foreground queue-wait p95 by 3.1$\times$ over FIFO, lowers p95
time-to-first-token (TTFT) by 1.55$\times$ versus non-preemptible training,
cuts post-publish first-hit prefill p99 by 25.6\% and cross-agent TTFT p99
by 21.9\%, and keeps background learning progressing under tight KV budgets.

\end{abstract}




\maketitle

\input{introduction}

\input{background}

\input{overview}

\input{unified_scheduler}

\input{version_aware_kv_cache_manager}

\input{multi_agent_runtime}

\input{evaluation}

\input{related_work}

\input{conclusion}

\bibliography{reference}

\end{document}

%% file: command.tex
\makeatletter
\newcommand{\removelatexerror}{\let\@latex@error\@gobble}
\makeatother

\newcommand{\eat}[1]{}

\AddToHook{env/figure/begin}{\vspace{-0.3em}}
\AddToHook{env/figure/end}{\vspace{-0.6em}}
\AddToHook{env/figure*/begin}{\vspace{-0.3em}}
\AddToHook{env/figure*/end}{\vspace{-0.6em}}

\newcommand{\reffig}[1]{Figure~\ref{fig:#1}}

\newcommand{\stitle}[1]{\vspace{1ex} \noindent{{\bf #1}}}

\usepackage{xspace}
\newcommand{\name}{LOCAL\xspace}

%% file: introduction.tex
\section{Introduction} \label{sec:intro}

Local LLM deployment is becoming practical on consumer hardware, driven
largely by privacy. Systems such as
MLC-LLM~\cite{mlc_llm} and ExecuTorch~\cite{executorch} target mobile and
edge backends, while PowerInfer~\cite{song2023powerinfer} and LLM in
Flash~\cite{alizadeh2024llmflash} reduce the computational and memory barriers to running
large language models locally.
Running models locally keeps private data, such as conversation history, personal
preferences, and tool interactions, on the user's device instead of sending it
to a remote service.

These local deployments are also beginning to move beyond one-shot inference
toward interactive agents that could improve from user feedback. An
\emph{on-device LLM agent} maintains session state, invokes tools, observes
outcomes, and revisits the same personal context across many turns~\cite{yao2023react,schick2023toolformer,wang2023voyager}. These
repeated interactions produce traces that are both private and valuable for
adaptation---user corrections, repeated preferences, failed tool calls, and
follow-up questions that reveal whether an earlier response helped~\cite{shinn2023reflexion,wang2026openclawrl}.
LoRA-style adapters~\cite{hu2022lora} make such adaptation feasible on a
single device: instead of fine-tuning the full model, only small low-rank
modules are trained, and different adapters can specialize the same base model
for different roles or preferences with minimal memory overhead. Sending
interaction traces to a remote trainer weakens the privacy motivation for local
deployment; keeping adaptation local preserves that motivation, but it means
that inference, feedback recovery, and adapter updates must share the same
device resources. The ideal is \emph{contiguous learning}: the agent adapts
from every interaction without interrupting user-facing inference, as though
learning and serving were a single unbroken process.

Neither existing inference runtimes nor existing RL training systems can support
contiguous on-device learning. Inference engines such as vLLM~\cite{kwon2023efficient} and
SGLang~\cite{zheng2024sglangefficientexecutionstructured}
are built around a \emph{stable-weight assumption}: loaded model parameters do
not change during serving, so KV-cache entries remain valid as long as the
visible token prefix matches. This assumption makes prefix caching safe and
scheduling simple, but it breaks continuity when an online training loop updates adapter
weights on the same device---cached KV tensors produced under an older adapter
version are no longer valid, and reusing them silently mixes hidden states
from different adapter states. Distributed RL systems such as
HybridFlow~\cite{sheng2024hybridflow}, OpenRLHF~\cite{hu2024openrlhf}, and
ReaL~\cite{mei2025real} adopt a \emph{resource-separation assumption}:
rollout, reward computation, reference scoring, and training can be placed on
separate accelerators and optimized for aggregate throughput. This assumption
also breaks continuity for on-device agents: a single consumer GPU may have to
respond to the user, run judge inference, train an adapter, and maintain KV state
simultaneously, with no separate cluster to absorb the load. The two assumptions that make existing systems work---stable
weights and separated resources---are precisely what prevent contiguous learning.

To enable contiguous learning for on-device LLM agents, we present \name, a
single-GPU runtime in which inference and adaptation share the same model
instance without interrupting each other. Realizing this requires addressing
three challenges.

\stitle{(1) Foreground--background contention on a single GPU.}
Foreground inference, judge inference, training steps, and cache refresh all
compete for the same GPU. A foreground-only policy preserves responsiveness
but starves background learning and cache maintenance; a background-heavy policy creates visible stalls.
Training is not cheaply preemptible at arbitrary points: safe interruption
boundaries occur around mini-steps and optimizer commits, so the runtime must
represent background work as bounded tasks that yield at these boundaries and
admit them only when the foreground path leaves a usable idle window.

\stitle{(2) KV-cache reuse under evolving adapters.}
KV caching is central to interactive throughput, but adapter updates break the
token-equality assumption used by ordinary prefix caches. Cached KV tensors
generated under adapter version~$v$ are unsafe to reuse after the adapter
commits version~$v{+}1$, even if the visible token prefix is identical.
Reusing such entries silently mixes hidden states from different adapter
states, while invalidating all stale entries destroys long-prefix reuse after
every update. The runtime therefore needs a cache contract that remains correct when adapter versions change, rather than relying on token prefix equality alone.

\stitle{(3) Memory management under constrained and contended VRAM.}
Consider a 24\,GB consumer GPU running a 7B-parameter model in FP16. The
weights alone occupy roughly 13.5\,GB; the remaining budget must accommodate
KV cache for long agent contexts, LoRA adapter parameters, judge and replay
batches, training activations, gradients, optimizer state, and checkpoint
staging. These objects have different lifetimes---KV cache persists across
turns, training activations exist only during a gradient step, and adapter
weights change at commit boundaries---but they compete for the same physical
memory. The pressure grows when multiple agents share one model instance:
several agents may carry long overlapping contexts, but each LoRA adapter may
need its own valid KV entries after adapter-specific updates, and a naive
policy that duplicates shared prefixes or retains stale entries can quickly
exhaust VRAM.

The root difficulty is that these three concerns are coupled: foreground
latency, adapter publication, and KV-cache validity cannot be managed by
independent subsystems without breaking learning continuity. Treating them independently leads to unbounded
foreground stalls, stale KV reuse, or memory exhaustion---all forms of
discontinuity. The key insight behind \name is that the three components responsible for these concerns---the scheduler, the model runtime, and the KV-cache manager---must operate over shared state. Adapter version, task priority, and cache validity are visible to all three so that scheduling decisions, execution, and cache maintenance remain mutually consistent and learning proceeds without interruption. This paper makes the following
contributions:


\stitle{(1)} We design a \emph{cooperative scheduler}
(\S\ref{sec:scheduler}) that centralizes GPU admission: it separates the
foreground inference queue from the background task queue, gives foreground
requests strict priority, and admits background judge, training, and
cache refresh tasks only as bounded chunks that yield at safe boundaries
when the foreground path leaves usable GPU windows.

\stitle{(2)} We design a \emph{version-aware KV-cache manager}
(\S\ref{sec:kv-cache}) that makes adapter version part of the cache key so
that stale coverage is rejected at lookup. The manager maintains hot-prefix
statistics to decide which entries are worth retaining, refreshing, or
offloading: when an adapter commits a new version, it issues stale-coverage prefill
instructions for hot prefixes; when the model runtime reports memory
pressure, it offloads colder entries to host memory according to the same
statistics. A radix-indexed prefix table separates logical validity from
physical residency.

\stitle{(3)} We design a \emph{multi-agent model runtime}
(\S\ref{sec:multi-agent}) that executes model inference and training tasks,
senses current memory pressure, and communicates offload requests to the
KV-cache manager. The runtime extends the cache contract with agent-scoped
KV identity---a cached entry is reusable only when its token prefix, context
namespace, logical adapter, and adapter version all match the request---and
adapter-scoped invalidation that limits the effect of training so that one
adapter publish does not invalidate KV owned by other adapters. Cross-agent
pre-prefill uses multi-agent communication structure to propose speculative
cache work without bypassing the scheduler.

We implement a prototype and evaluate it on a single consumer GPU with
7B-class models. Under mixed foreground inference, judge inference, training,
and cache-maintenance work, \name reduces foreground queue p95 and p99 by
3.1$\times$ and 3.4$\times$ over FIFO while completing 34 adapter training
commits. Cooperative interruption makes p95 TTFT
1.55$\times$ lower than non-preemptible training and reduces accumulated
training blocking by 2.6$\times$. Compared with a vLLM full-LoRA path, \name
lowers p95 queue wait by 5.9$\times$--12.8$\times$ across admission policies.
Its cache mechanisms lower post-publish first-hit prefill p99 by 25.6\%,
reduce cross-agent TTFT p99 by 21.9\%, and keep the runtime operating while
peak GPU memory ranges from 21.6\,GB to 23.5\,GB.

%% file: background.tex
\section{Background and Motivation} \label{sec:background}

\subsection{On-Device LLM Agents}

On-device LLM deployment keeps private data local, reduces dependence on
network services, and can respond to users when connectivity is unavailable.
Recent work has made this setting increasingly practical from several
directions: deployment runtimes and compiler stacks target mobile and edge
backends~\cite{mlc_llm,executorch,litert_lm}, while inference engines exploit
consumer-grade CPU/GPU or flash/DRAM memory
hierarchies~\cite{song2023powerinfer,alizadeh2024llmflash}.
An \emph{on-device LLM agent} is a local application-level control loop around
such a model. It maintains session state, receives user instructions, invokes
tools, observes tool or environment outcomes, and uses later turns to decide
what to do next. Unlike one-shot local inference, the agent repeatedly revisits
the same personal context on the same device.

This setting exposes private traces that are useful for adaptation: user
corrections, repeated preferences, failed tool results, abandoned plans, and
follow-up questions that clarify whether an earlier response helped. Sending
these traces to a remote trainer weakens the privacy motivation for local
deployment. Keeping adaptation local preserves that motivation, but it also
means that foreground inference, judge inference, and model updates must share the same
device resources.

Agent interaction is therefore not a stateless prompt-response call. A typical
agent session is multi-turn: the model proposes an action, invokes a tool or
environment operation, observes the result, and uses that observation in later
turns. ReAct formalizes this interleaving of reasoning, acting, and observation
for language agents~\cite{yao2023react}; related agent work extends the pattern
to API/tool use, verbal feedback, and embodied environments~\cite{schick2023toolformer,shinn2023reflexion,wang2023voyager}.
As applications become more complex, multi-agent frameworks such as CAMEL,
AutoGen, and MetaGPT further decompose applications into conversations or
workflows among specialized agents~\cite{li2023camel,wu2023autogen,hong2024metagpt}.
This structure is useful for local assistants because a planner, tool agent,
critic, judge, or coordinator may share task context while applying different
prompts, tools, or adaptation policies.

\subsection{Online RL for On-Device Agents}

On-device agents need to learn from the traces their interactions produce.
Online reinforcement learning (RL) provides a natural framework: a
\emph{policy} generates responses, a \emph{reward source} evaluates them, and
an \emph{optimizer} updates the policy. In the agent setting, the reward
comes from the next state after an action---user replies, tool outputs,
terminal states, and GUI state changes all carry information about whether the
previous response helped. We adopt the term \emph{next-state feedback} for
this signal, following recent work that formalizes it for personal
agents~\cite{wang2026openclawrl}. The signal is useful but delayed and noisy:
a local runtime must record enough metadata on the foreground path and later
recover rewards, build replay samples, and update adapters without blocking
the next foreground inference.

Two technical mechanisms make on-device RL both feasible and challenging.

\stitle{LoRA adapters} make training feasible within a single device's memory
budget. Instead of fine-tuning the full model, LoRA adds small trainable
low-rank modules while sharing a frozen base model~\cite{hu2022lora}; recent
work has shown that on-device training with LoRA is practical when memory
pressure is managed carefully~\cite{gao2025mobizo}. For agents, different
adapters can represent different roles, preferences, or task types, and a
runtime must route each request to the correct adapter, publish adapter
updates atomically, and avoid using partially updated weights for inference.
However, adapter updates also invalidate previously valid inference state, as
we explain next.

\stitle{KV cache and prefix caching} make multi-turn inference efficient but
break when adapters change. During inference, the model stores key and value
tensors (KV cache) for previous tokens so that later decode steps attend to
cached tensors instead of recomputing the full history. Prefix caching further
reuses KV tensors when a new request shares a token prefix with a previous
request: vLLM shows that paged KV-cache management is important for
high-throughput inference~\cite{kwon2023efficient}, while SGLang shows that
prefix reuse is important for efficient execution of structured LLM
programs~\cite{zheng2024sglangefficientexecutionstructured}. These
optimizations rely on an implicit contract: the model weights that produced
the cached tensors have not changed. When an adapter is updated from version
$v$ to $v{+}1$, cached KV tensors produced under version $v$ are no longer
valid---reusing them silently mixes hidden states from different adapter
states. On a device where inference and training coexist, this contract is
continuously broken.

\subsection{Co-Locating Inference and Training}

\begin{figure}[t]
    \centering
    \subfigure[Distributed RL runtime]{
        \includegraphics[height=0.44\columnwidth]{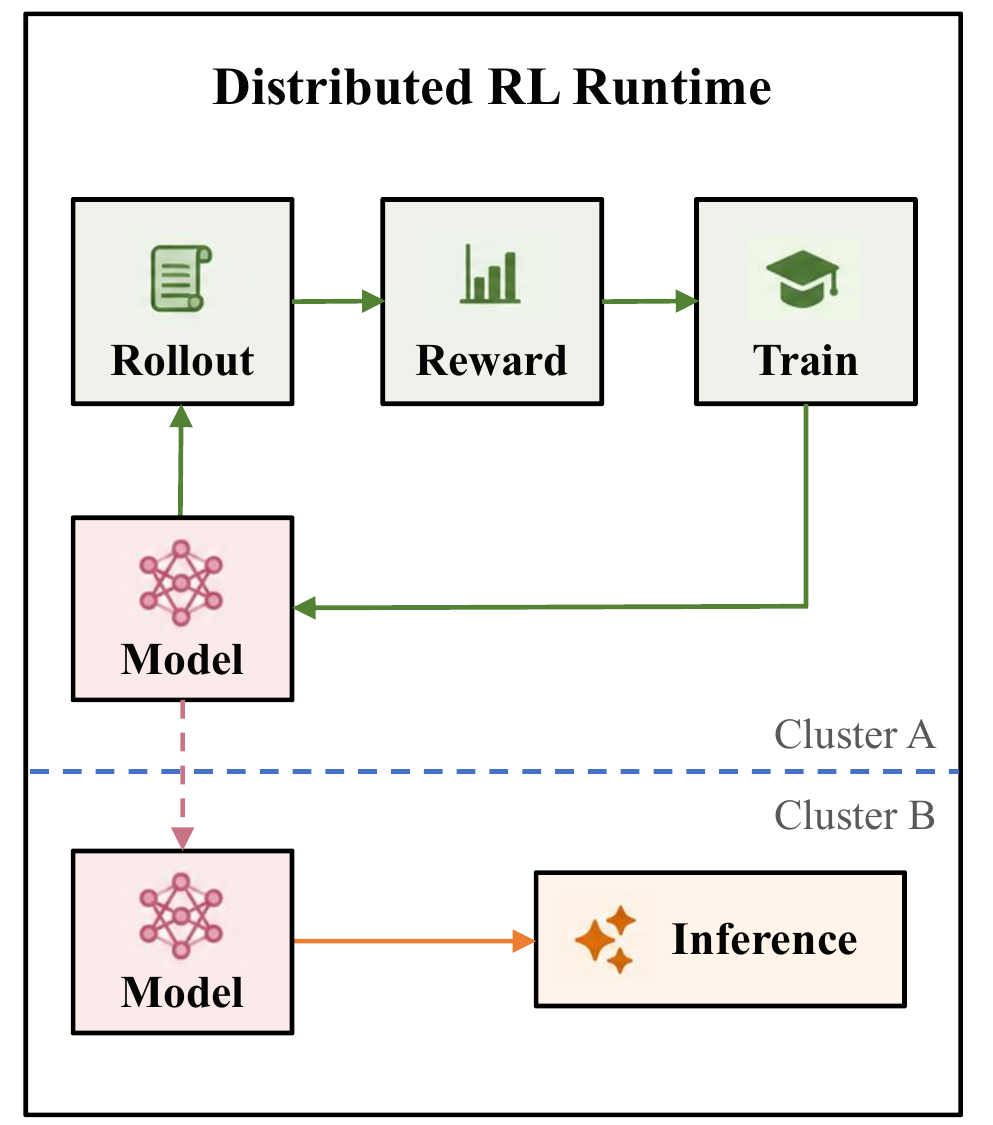}
        \label{fig:rl_comparison_a}
    }
    \hfill
    \subfigure[On-device agent runtime]{
        \includegraphics[height=0.44\columnwidth]{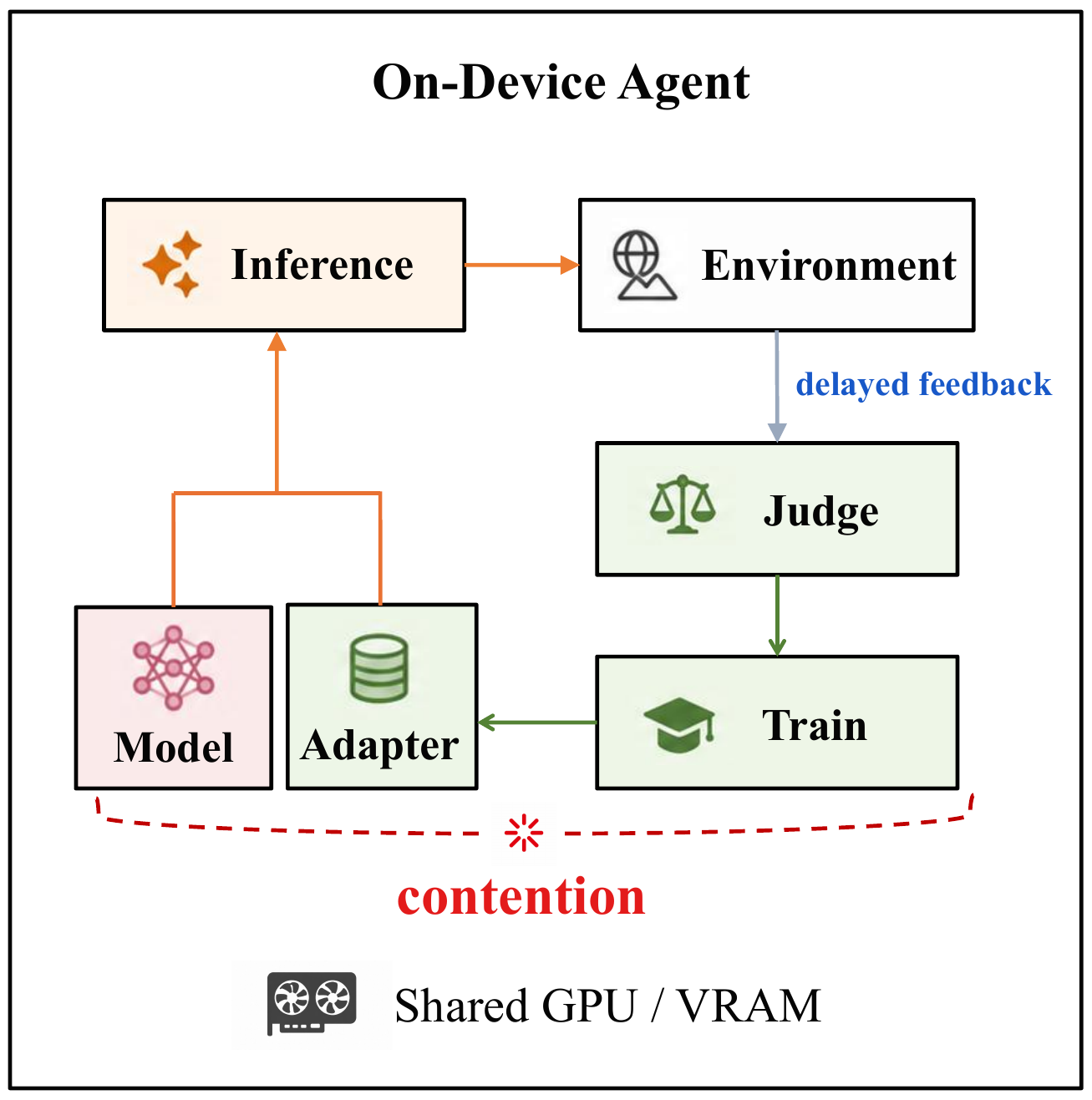}
        \label{fig:rl_comparison_b}
    }
    \caption{Resource coupling in on-device learning.}
    \label{fig:rl_comparison}
\end{figure}

Existing local and production inference systems are built around a relatively
stable inference state. Local and edge runtimes such as MLC-LLM and ExecuTorch
support device-specific deployment and execution, while inference systems such
as vLLM and SGLang optimize batching, KV-cache management, and prefix reuse for
loaded weights~\cite{mlc_llm,executorch,kwon2023efficient,zheng2024sglangefficientexecutionstructured}.
In both cases, the common contract is that weights used for inference are not continuously
mutated by an online training loop on the same device. Existing RL systems make
a different resource assumption: rollout workers, reward workers, and training
workers can run on different devices and optimize aggregate training
throughput~\cite{sheng2024hybridflow,hu2024openrlhf,mei2025real}. Recent
RL/RFT frameworks further broaden the design space with asynchronous rollout,
flexible data generation, and agent-environment interaction
support~\cite{slime_github,fu2025areal,pan2025trinityrft}. On-device agent
training breaks both assumptions, as shown in
\reffig{rl_comparison}. The inference state
evolves because adapters are updated from local feedback, and the rollout,
reward, training, and cache-maintenance work all contend for one local model
instance.

This coexistence requirement is the core problem. The runtime cannot treat inference, training, and cache management as independent subsystems, because adapter updates affect inference-time KV validity, cache retention affects training memory, and delayed feedback requires attribution to the agent and adapter that produced a response.

\subsection{Challenges of On-Device Inference-Training Coexistence}

\begin{figure}[t]
    \centering
    \includegraphics[width=\columnwidth]{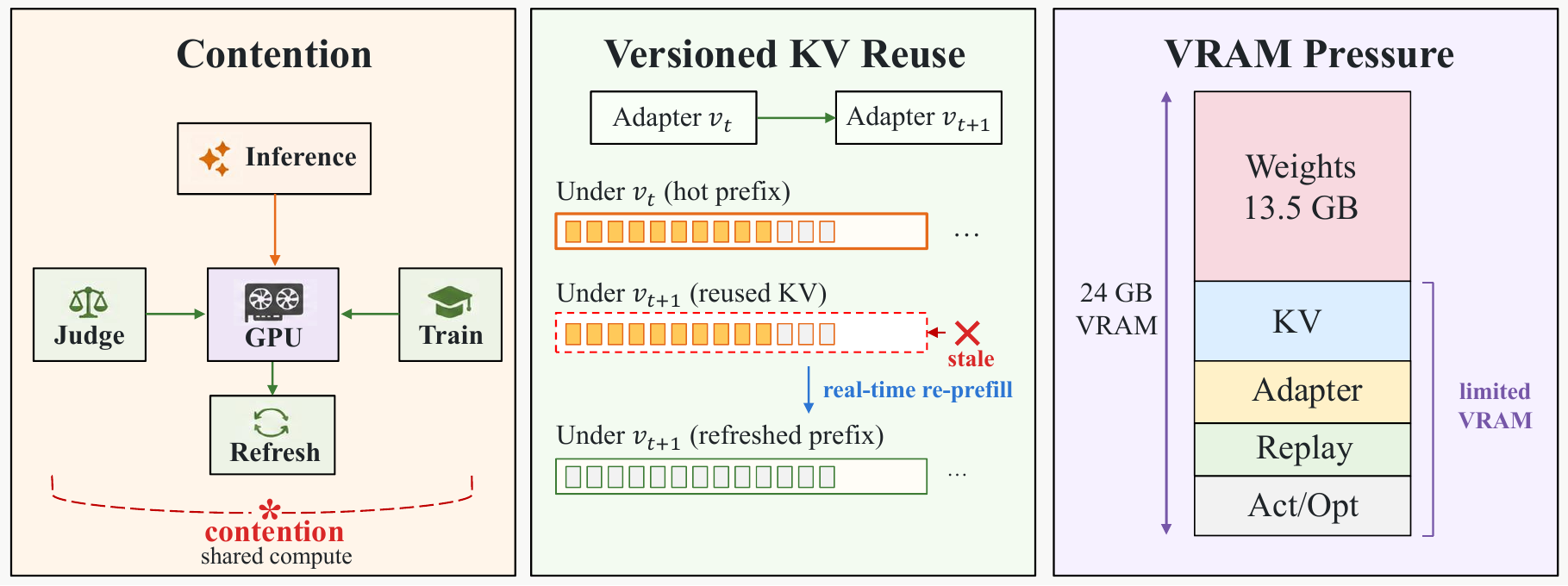}
    \caption{Runtime challenges for on-device RL.}
    \label{fig:challenges_on_device_training}
\end{figure}

The hardware envelope is close to a consumer device: one model instance, one
consumer-grade GPU, and no separate accelerator dedicated to reward modeling
or training. Consider a 24\,GB consumer GPU running a 7B model in
FP16/BF16. The weights alone occupy roughly 13.5\,GB. The remaining gigabytes
must accommodate KV cache for long context, LoRA adapters, judge and replay
batches, training activations, gradients, optimizer state, and checkpoint
staging. \reffig{challenges_on_device_training} summarizes the coupled
runtime challenges in this envelope.

\stitle{(1) Foreground--background contention on a single GPU.}
Foreground inference, judge inference, training steps, and cache refresh
all compete for the same GPU. A foreground-only policy
preserves responsiveness but starves judging and training under frequent
interaction; a background-heavy policy advances training but creates visible
stalls when training steps allocate activations and optimizer state. Worse, training is not
cheaply preemptible at arbitrary points: safe interruption boundaries occur
around mini-steps, optimizer commits, or adapter publication. The
runtime therefore needs a single admission point that represents background
work as bounded tasks, exposes safe interruption boundaries, and admits them
only when the foreground path leaves a usable idle window.

\stitle{(2) KV-cache reuse under evolving adapters.}
KV caching is central to interactive throughput, but adapter updates break the
token-equality assumption used by ordinary prefix caches. A cached entry
generated under adapter version $v$ is not generally valid after the adapter
commits version $v{+}1$, even if the visible token prefix is identical.
Reusing such an entry silently mixes computations from different adapter
states, while invalidating all stale entries destroys long-prefix reuse after
every update. The runtime therefore needs a cache contract in which adapter
identity and version are part of the cache key, stale coverage is rejected at
lookup, and hot prefixes can be refreshed under the new version outside the
foreground critical path.

\stitle{(3) Memory management under constrained and contended VRAM.}
After loading a 7B model in FP16/BF16, a 24\,GB local GPU has only a small
remaining budget for mutable runtime state. This budget must be shared by KV
cache, judge prompts, replay batches, LoRA parameters, training activations,
gradients, optimizer state, and checkpoint staging. These objects have
different lifetimes and correctness constraints, but they compete for the same
physical memory. KV cache is the largest inference state that grows with interaction length.
Retaining long prefixes reduces time-to-first-token, but reduces the memory
available for judging, training, and future foreground requests. Evicting
aggressively preserves headroom for training, but destroys prefix reuse. With
multiple adapters, KV reuse is further scoped by adapter identity and version,
not only by visible tokens. Multi-agent execution makes the pressure worse:
several agents may carry long overlapping contexts, but each LoRA adapter may
need its own valid KV entries after adapter-specific updates. A naive policy
that duplicates shared prefixes per agent or retains stale entries can quickly
exhaust VRAM. Therefore, memory management cannot be a local cache policy
alone: shared metadata is needed about prefix utility, residency, adapter
provenance, version, workspace demand, and memory pressure in order to
coordinate retention, offload, refresh, and training admission.

These challenges motivate three runtime mechanisms developed in the rest of
the paper: a cooperative scheduler that centralizes GPU admission and bounds
background work (\S\ref{sec:scheduler}), a version-aware KV-cache manager
that maintains cache correctness and hot-prefix-driven maintenance
(\S\ref{sec:kv-cache}), and a multi-agent model runtime that executes
inference and training, senses memory pressure, and extends the cache contract
with agent-scoped identity (\S\ref{sec:multi-agent}).





%% file: overview.tex
\section{System Overview} \label{sec:overview}

The design of \name is grounded in one insight.

\stitle{Key insight.} Foreground latency, adapter publication, and KV-cache
validity cannot be managed independently without breaking learning continuity.
Background online RL changes adapter parameters, which invalidates KV tensors
produced by older adapter versions even when the visible token prefix is
unchanged. \name therefore treats adapter version, task priority, and cache
validity as first-class runtime state, so that the three
components responsible for these concerns---the scheduler, the model runtime,
and the KV-cache manager---operate over the same state and keep learning
uninterrupted.

\begin{figure}[t]
    \centering
    \includegraphics[width=\columnwidth]{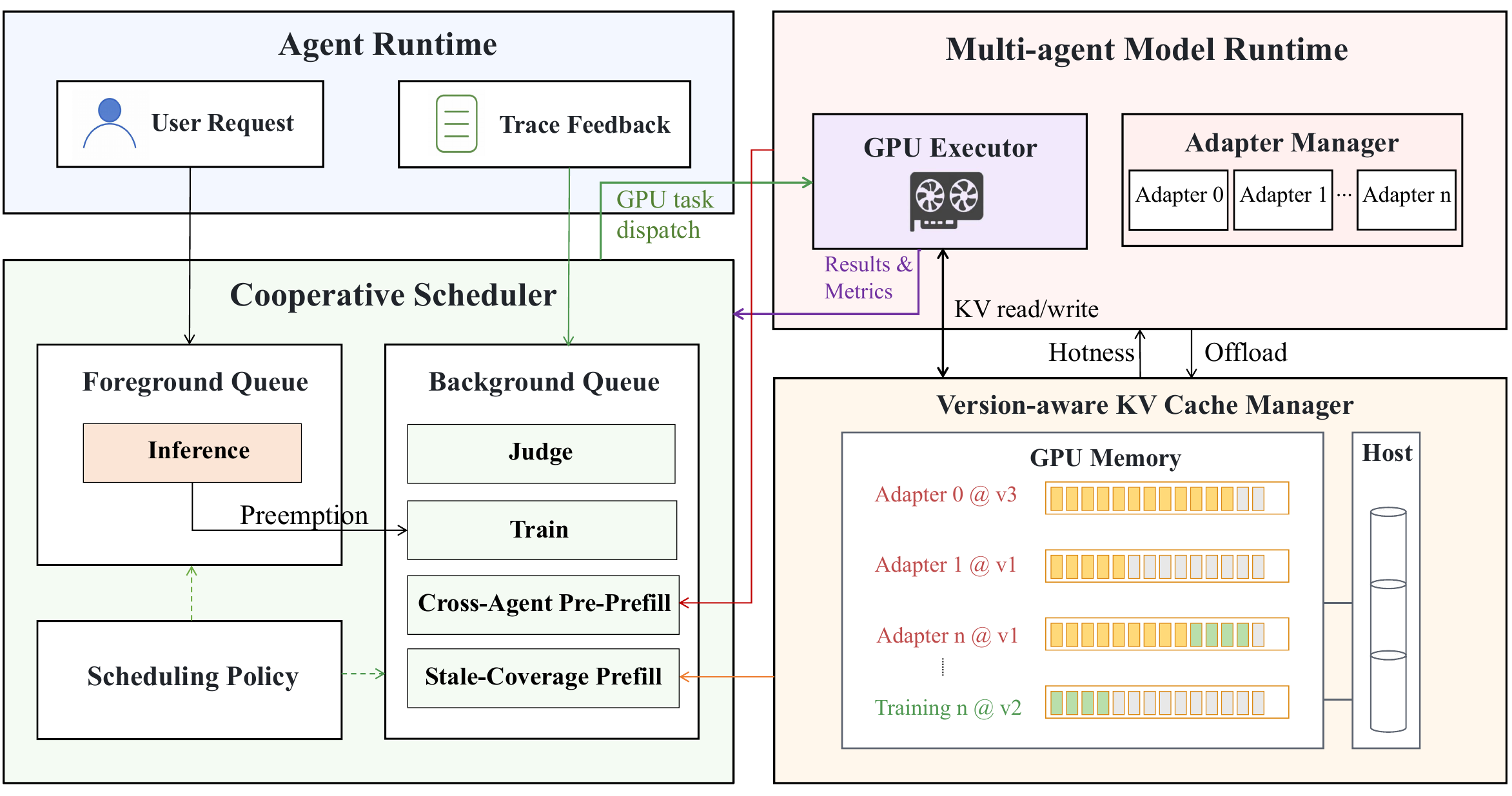}
    \caption{\name architecture.}
    \label{fig:overview}
\end{figure}

Figure~\ref{fig:overview} shows how \name realizes this insight. The runtime
sits between a local agent API and a single model instance and is organized
into three components that share adapter version, task priority, and cache
validity as common state: a \emph{cooperative scheduler} that is the only GPU
admission point, a \emph{version-aware KV-cache manager} that owns cache
validity and hot-prefix metadata, and a \emph{multi-agent model runtime} that
executes inference and training, senses memory pressure, and communicates with
the other two components.

\emph{Cooperative scheduler}~(\S\ref{sec:scheduler}). The scheduler
separates the foreground inference queue from the background task queue and
admits only one active GPU task at a time. Foreground requests retain strict
priority; background judge inference, training, and cache-maintenance tasks
run only at safe yield or commit boundaries when the foreground path leaves
usable GPU windows.

\stitle{Version-aware KV-cache manager}~(\S\ref{sec:kv-cache}). The KV-cache
manager makes adapter version part of the cache key so that stale coverage is
rejected at lookup. It maintains hot-prefix statistics to decide which entries
are worth retaining, refreshing, or offloading. When an adapter commits a new
version, the manager issues re-prefill instructions for hot prefixes as
scheduler-visible background tasks. When the model runtime reports memory
pressure, the manager offloads colder entries to host memory according to the
same statistics. A radix-indexed prefix table separates logical validity from
physical residency.

\stitle{Multi-agent model runtime}~(\S\ref{sec:multi-agent}). The model runtime
executes model inference and training tasks, senses current memory pressure,
and communicates offload requests to the KV-cache manager. When multiple
agents share one model instance, the runtime preserves per-agent identity,
adapter binding, and KV provenance across inference, judging, replay, and
adaptation. Adapter-scoped invalidation limits the effect of training: when one
adapter commits a new version, only entries owned by that adapter become stale;
entries owned by other adapters remain valid. Cross-agent pre-prefill uses
multi-agent communication structure to propose speculative cache work without
bypassing the scheduler.

The current prototype targets one local GPU, one base model instance, and
LoRA-style adapter updates. It does not assume a remote rollout, reward, or
training cluster. The rest of the paper details the three components inside
this boundary.

%% file: unified_scheduler.tex
\section{Cooperative Scheduler} \label{sec:scheduler}

This section describes \name's cooperative scheduler, which bounds foreground
interference when serving, judging, training, and stale-coverage prefill share
one GPU. Other runtime components may prepare work, but only the scheduler may
admit GPU-bound model execution. This invariant gives \name one place to reason
about task priority, estimated cost, memory pressure, adapter visibility, and
safe yield boundaries. The scheduler represents heterogeneous model operations
with a common task interface, gives foreground inference strict admission
priority, and admits background work only as bounded chunks whose memory demand
and non-yielding duration are visible before execution.

\subsection{Task Model}

Every GPU-bound model operation is represented as a task. A foreground
inference task carries the request, agent identity, adapter binding, and cache
context needed to return a user-visible response. A judge task carries batched
inference inputs, a training task carries a prepared replay batch and safe yield
or commit boundaries, and a stale-coverage prefill task carries a prefix chunk,
target adapter version, cache priority, and refresh state. Preparation alone
does not grant GPU access: ready tasks enter either the foreground inference
queue or the background task queue, and the scheduler invokes the GPU
single-task executor only after admission.

Tasks expose the metadata needed to make heterogeneous work comparable. The
scheduler does not require one global cost model for the whole runtime; it only
needs local estimates sufficient to decide whether a task can fit the current
time and memory window. For a task $x$, the scheduler records
\[
    d(x) = \big(c_x, p_x, a_x, \hat{t}(x), \hat{m}(x),
    \hat{y}(x), u(x)\big),
\]
where $c_x$ is the task class, $p_x$ is its hard priority, $a_x$ is the required
adapter state, $\hat{t}(x)$ is the estimated admitted-chunk duration,
$\hat{m}(x)$ is the estimated peak memory or KV-page demand, $\hat{y}(x)$ is
the longest non-yielding interval before the next safe boundary, and $u(x)$ is
the task-specific utility used to rank otherwise admissible background work.

Duration estimates are calibrated online from completed chunks. For each
measured quantity $z$ such as judge duration, training-chunk duration, or
prefill seconds per token, \name maintains an exponential moving average
\[
    E_z^{(i)} =
    \begin{cases}
        v_i, & E_z^{(i-1)} \text{ is unset},\\
        (1-\alpha_z)E_z^{(i-1)} + \alpha_z v_i, & \text{otherwise},
    \end{cases}
\]
where $v_i$ is the most recent observation. Judge and training estimates use
class-specific duration EMAs. Prefill additionally records a per-token rate
$\hat{\rho}_{b(L)}$ for prefix-length bucket $b(L)$, giving
\[
    \hat{t}_{\mathrm{pf}}(\Delta,L)
      = \hat{\rho}_{b(L)} \Delta,\qquad
    b(L) \in \{512,1024,2048,\ldots,32768\}.
\]
The prototype also caps prefill chunks by configured chunk size, available KV
pages, and foreground reserve. Before a judge or training duration sample
exists, that task family is admitted only in an explicit turn-finished
background window.

\subsection{Foreground Priority and Cooperative Training}

Foreground inference is always considered before new background work because
it is the only task type on the user-visible latency path. When the foreground
inference queue is non-empty, the scheduler dispatches foreground inference
and does not start new judge, training, or stale-coverage prefill tasks. This
rule is deliberately simple: it makes the latency policy easy to audit and gives
the agent runtime a stable contract.

The remaining problem is already-running background work. \name does not
assume arbitrary preemption of GPU kernels, model forward passes, backward
passes, or optimizer updates. Instead, training follows a cooperative
interruption contract. When a foreground request arrives while training is
running, the scheduler issues an interruption signal. The training task observes
this signal only at safe boundaries inside the step, after a non-yielding phase
has completed and before the next phase begins. Once the signal is observed,
training stops at that boundary and returns control to the scheduler, allowing
foreground inference to be admitted.

This contract bounds foreground waiting time without requiring unsafe GPU
preemption. A foreground request cannot stop a model kernel or optimizer update
in the middle of execution; it can only prevent the training task from entering
the next non-yielding phase. The blocking time seen by the foreground path is
therefore bounded by the remaining time to the next safe training boundary,
rather than by the duration of an entire training procedure.

An interrupted training task never exposes partial adapter state to foreground
inference. Foreground inference binds only the last committed serving adapter;
in-progress training state remains private until the adapter manager publishes a
staged version at a controlled publication point.

\begin{figure}[t]
    \centering
    \includegraphics[width=\columnwidth]{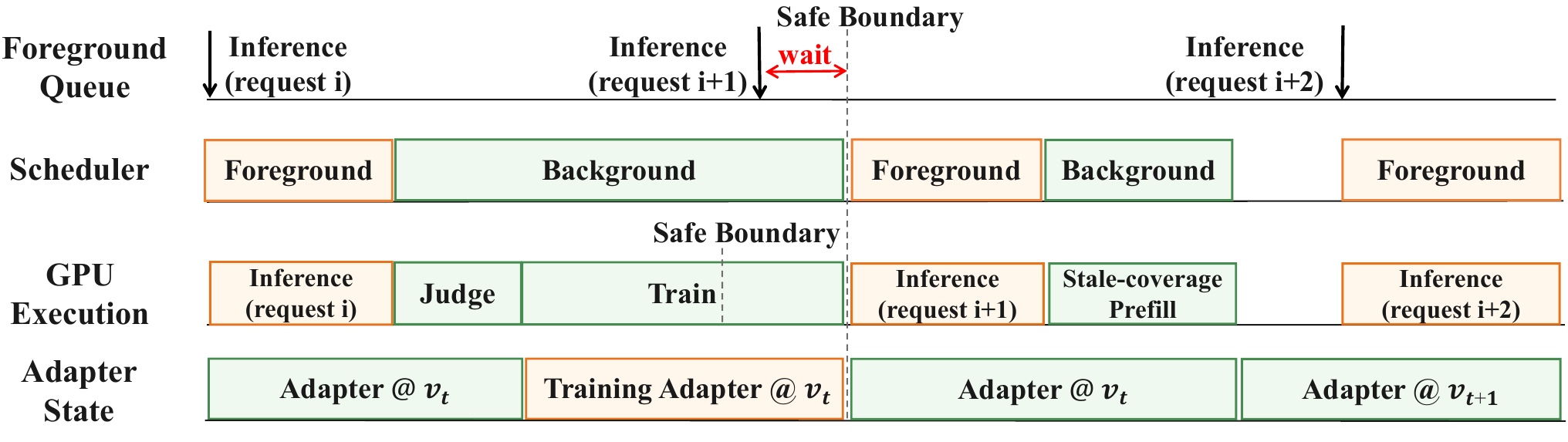}
    \caption{Bounded background admission.}
    \label{fig:scheduling-timeline}
\end{figure}

Figure~\ref{fig:scheduling-timeline} illustrates this contract. A foreground
request does not interrupt arbitrary GPU execution. Instead, it prevents new
background admission and takes over at the next safe yield boundary of the
currently running background chunk.

\subsection{Timeslice-Aware Background Admission}

When no foreground task is ready, the scheduler admits background work
opportunistically. For each ready background task, the scheduling policy
considers estimated duration, estimated peak memory, next-yield interval, task
priority, and replay or cache utility. A short judge batch may fit where a
training chunk does not; a bounded stale-coverage prefill chunk may fit where a full
prefix rebuild would not. The scheduler therefore treats background progress as
a sequence of small, schedulable units rather than as a fixed background loop.

Let $M_{\mathrm{free}}$ denote currently available GPU memory,
$M_{\mathrm{reserve}}$ denote the reserve kept for unexpected foreground work,
and $W_{\mathrm{bg}}$ denote the conservative background window selected by the
runtime. A background task $x$ is admissible only if
\[
    \hat{m}(x) + M_{\mathrm{reserve}} \le M_{\mathrm{free}}
    \quad \text{and} \quad
    \hat{y}(x) + \delta \le W_{\mathrm{bg}} ,
\]
where $\delta$ is a conservative latency margin. When the agent runtime exposes
a tool or subtask deadline $D_{\mathrm{tool}}$, the policy also checks the
remaining window
\[
    R = D_{\mathrm{tool}} - t_{\mathrm{now}},
    \qquad
    \hat{t}(x) + \delta \le R .
\]
Among admissible tasks, the scheduler prefers the task with the highest
foreground-safe utility, such as urgent reward recovery, a training update close
to publication, or a stale-coverage prefill chunk for a hot prefix. If no task satisfies
the memory and yield constraints, the scheduler leaves the GPU idle briefly or
asks the KV cache manager to create room by offloading cold spans.

This policy does not rely on perfect prediction of the next user request.
Foreground interference is controlled by bounding the non-yielding portion of
each admitted background task. If a duration estimate is too small, the excess
cost is limited to the current chunk; once a foreground request arrives, no new
background task is admitted. If a memory estimate is too small, page or allocator
checks reject the task, or the runtime offloads lower-priority cache state before
retrying. A failed training chunk does not publish a staged adapter, and a
failed prefill chunk does not create reusable KV coverage.

Timeslice admission also centralizes memory backpressure. Under high memory
pressure, the scheduler can reduce cross-agent pre-prefill, avoid starting a
training chunk, or prefer judge inference over training until offload has
created sufficient headroom. This keeps memory decisions in the shared runtime
policy rather than scattering them across unrelated background pipelines.

Overall, the cooperative scheduler converts on-device learning from a set of competing
threads into a single GPU-admission problem. Foreground inference keeps strict
priority, background learning progresses in bounded chunks, training publishes
adapter updates only at safe boundaries, and stale-coverage prefill remains correct
because every refreshed span is tied to an explicit adapter version.

%% file: version_aware_kv_cache_manager.tex
\section{Version-Aware KV Cache Manager} \label{sec:kv-manager}\label{sec:kv-cache}

The version-aware KV cache manager owns the reusable inference state shared by
foreground inference, online adaptation, and multi-agent cache preparation. It
does not admit GPU work directly. Instead, it maintains cache identity,
hot-prefix statistics, residency state, and adapter-version staleness, then
exports bounded cache-maintenance tasks and offload candidates to the scheduler.
The manager provides four mechanisms: agent-scoped KV identity, hot-prefix
statistics, stale-coverage prefill for staged adapter versions, and
priority-based retention/offload under memory pressure.

\subsection{Agent-Scoped KV Identity}

Multi-agent workflows create repeated prefixes from task descriptions, shared
environment observations, tool histories, role prompts, and mailbox state. These
repetitions make prefix reuse valuable, especially on a single GPU where long
prefill windows compete with foreground latency and background learning. However,
the same visible prefix may be executed under different roles, different adapters, or different adapter versions. Reusing KV tensors across incompatible
contexts would mix hidden states produced under different computation states.

\name therefore separates \emph{validity keys} from \emph{provenance metadata}.
A KV entry is reusable only when its validity key matches the request:
\[
    \kappa = (\textit{token span},\ \textit{context namespace},\
              \textit{adapter},\ \textit{adapter version}) .
\]
The token span identifies the textual prefix covered by the KV tensors. The
context namespace distinguishes role- or session-specific state that is not
otherwise captured by the token span and adapter binding. The adapter
identifies the LoRA slot selected for the request, and the adapter version
identifies the committed serving state that produced the KV tensors.

Agent identity is recorded as provenance rather than always being a hard cache
key. If two agents share the same serialized prefix, context namespace, adapter,
and adapter version, their KV coverage may be shared even if the producer agents
differ. If agent identity changes the role prompt, private memory, mailbox view,
adapter routing, or communication context, that difference is reflected in the
context namespace or adapter binding. The cache manager stores these records in a
radix-indexed prefix table so shared token paths need not be duplicated for every
agent or adapter state.

\subsection{Hot-Prefix Statistics}

Correct invalidation prevents stale reuse, but it does not decide which KV spans
should remain resident, which stale spans should be refreshed, or which missing
spans should be materialized before they are requested. \name uses hotness as a
compact ranking signal for these cache-maintenance decisions. Instead of
modeling hotness as a collection of low-level implementation counters, the cache
manager decomposes KV demand into two factors: \emph{prefix hotness} and
\emph{adapter hotness}. A KV record is valuable only when both its token prefix
is likely to be reused and the adapter version that produced the hidden states
is likely to be selected again.

\begin{figure}[t]
    \centering
    \includegraphics[width=\columnwidth]{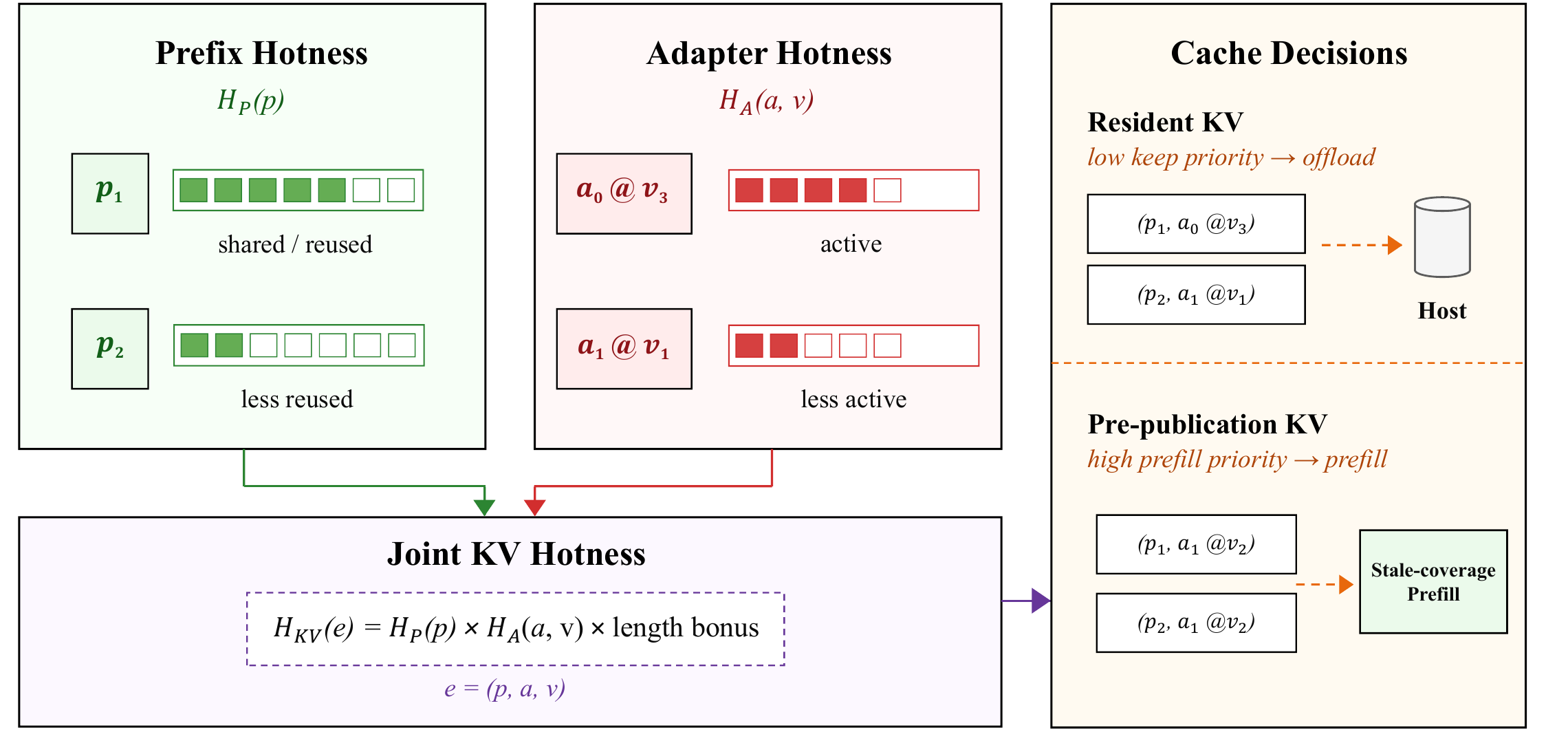}
    \caption{Version-aware KV management.}
    \label{fig:kv-manager}
\end{figure}

Let $p$ denote a radix-tree prefix and let $\nu=(a,v)$ denote an adapter and its
adapter version. The manager maintains a prefix heat $H_P(p)$ and an adapter
heat $H_A(\nu)$. Prefix heat is adapter-independent: it captures whether a token
span is repeatedly reached by foreground requests, shared among agents, or
predicted by the multi-agent runtime to be consumed by downstream turns. Adapter
heat is prefix-independent: it captures whether an adapter version is actively
serving traffic, routed by recent requests, or staged for imminent publication.
Both quantities are maintained as decayed statistics:
\[
    H_P(p) \leftarrow
    \lambda_P H_P(p) + (1-\lambda_P)
    \left(
        u_P(p) + \beta d_P(p) + \chi s_P(p)
    \right),
\]
\[
    H_A(\nu) \leftarrow
    \lambda_A H_A(\nu) + (1-\lambda_A)
    \left(
        u_A(\nu) + \eta d_A(\nu) + \zeta g_A(\nu)
    \right).
\]
Here $u_P(p)$ and $u_A(\nu)$ are observed-use signals from the current request,
$d_P(p)$ and $d_A(\nu)$ are predicted downstream demand signals, $s_P(p)$
captures cross-agent sharing of the prefix, and $g_A(\nu)$ marks adapter
versions that are staged or expected to become visible soon. The coefficients
only tune the relative importance of observed reuse, predicted reuse, sharing,
and staged-adapter urgency; the policy does not rely on an exact performance
model.

For a KV coverage record $e=(p,\nu)$, \name combines the two heat values into a
KV hotness score:
\[
    H_{\mathrm{KV}}(e) =
    \left(\epsilon + H_P(p)\right)
    \left(\epsilon + H_A(\nu)\right)
    \left(
        1+\alpha_L
        \frac{\min(L_e,L_{\max})}{L_{\max}}
    \right),
\]
where $L_e$ is the covered prefix length and $\epsilon>0$ prevents one dimension
from completely suppressing the other. The multiplicative form is intentional:
a globally popular adapter should not protect an irrelevant prefix, and a hot
prefix should not be retained under an adapter version that is unlikely to be
used. The length term favors longer reusable spans because they save more
prefill work, but caps the benefit to avoid letting very long prefixes dominate
the cache.

The hotness score is then converted into a retention priority for resident KV.
The cache manager ranks how valuable the object is to keep:
\[
    \Pi_{\mathrm{keep}}(e) =
    \frac{
        H_{\mathrm{KV}}(e)\cdot \widehat{C}_{\mathrm{recompute}}(e)
    }{
        \max(1,B_e)
    },
\]
where $B_e$ is the resident memory footprint and
$\widehat{C}_{\mathrm{recompute}}(e)$ estimates the cost of reconstructing the
coverage. Lower $\Pi_{\mathrm{keep}}$ means the resident KV is colder relative
to its memory cost and is therefore a better eviction candidate. For missing or
stale coverage, this section only records that the span may be useful to
materialize later; downstream mechanisms combine this heat with their own
demand signals.

At request time, the cache manager refreshes heat only along matched radix paths
and relevant adapter versions, avoiding a global scan. The updated records expose
three decisions: hot resident spans are protected from eviction, hot stale spans
become stale-coverage prefill candidates, and hot missing spans provide candidate
inputs for cross-agent pre-prefill. Spans with low joint prefix--adapter heat are
released first under memory pressure.

\subsection{Scheduler-Visible Stale-Coverage Prefill}

Stale-coverage prefill is the pre-publication cache warmup performed after
training produces a new adapter version but before that version becomes visible
to foreground inference. The adapter manager first places the newly trained
adapter in a staged state. While the old adapter continues to serve user
requests, the KV cache manager selects hot prefixes, creates bounded prefill
chunks for the staged adapter version, and submits those chunks to the scheduler
as ordinary background work.

The purpose of this mechanism is to reduce the time-to-first-token after an
adapter update. Without stale-coverage prefill, the first foreground requests
served by the new adapter would encounter cold hot-path coverage: although the
runtime may already have KV entries for the same token prefixes under the old
adapter, those entries cannot be reused because KV states are adapter-version
dependent. The foreground path would therefore need to rebuild the hot prefix
under the new adapter before decoding, increasing post-update TTFT. By computing
the hot-path KV coverage before the adapter is published, \name shifts this
cost into foreground-idle GPU windows and preserves the low-latency path after
the update becomes visible.

The scheduler treats each stale-coverage prefill chunk as explicit GPU work
rather than as a hidden cache optimization. A chunk carries the target staged
adapter version, the prefix span to prefill, its cache priority, estimated memory
demand, and the next safe yield boundary. The scheduler admits it only when the
foreground path has no waiting request and the chunk fits the current time and memory budget.
Thus, stale-coverage prefill competes with judge and training work under the
same background-admission policy, and it cannot bypass foreground priority.

This design separates adapter readiness from adapter visibility. During
stale-coverage prefill, foreground inference continues to bind the last
committed serving adapter, and KV generated for the staged adapter is kept
private until publication. If publication happens before all hot prefixes have
been warmed, the new adapter reuses only prefetched spans that match its version
and falls back to ordinary prefill for the remaining spans. Thus,
stale-coverage prefill trades idle GPU windows for lower post-update TTFT
without exposing partially updated adapter state or stale KV to the foreground
path.

Stale-coverage prefill is distinct from cross-agent pre-prefill
(\S\ref{sec:multi-agent}): both are background cache-maintenance tasks, but the
former is tied to adapter-version publication, while the latter is tied to
anticipated context sharing.

\subsection{Priority-Based Retention and Offload}

When the scheduler or multi-agent runtime requests memory headroom, the KV cache
manager chooses low-value resident spans according to $\Pi_{\mathrm{keep}}$.
The offload order is
\[
    \text{offload order} =
    \operatorname{sort}_{\mathrm{asc}}
    \big(\Pi_{\mathrm{keep}}(e), -B_e\big).
\]
Thus, resident KV with weak joint prefix--adapter heat is released first, and
ties release larger objects first. Active adapter weights and the active
request's KV state are not eligible offload candidates.
To satisfy a target release $R$, the manager filters resident records to
offload-eligible entries, applies this ordering, and releases the shortest
prefix of the ordered list whose total footprint reaches $R$.

This policy retains shared hot prefixes only under adapter versions likely to be
used, avoids protecting stale KV merely because the prefix was recently accessed,
and ranks background materialization with the same joint heat signal used for
retention.

%% file: multi_agent_runtime.tex
\section{Multi-Agent Runtime} \label{sec:multi-agent}

The multi-agent runtime sits above the cooperative scheduler
(\S\ref{sec:scheduler}) and the version-aware KV cache manager
(\S\ref{sec:kv-manager}). It turns agent handles, communication state, and
delayed feedback context into scheduler-visible work and cache-manager requests
when several local agents, adapters, and online adapter updates share one model
instance. KV validity, hotness, stale-coverage prefill, and concrete offload
selection remain owned by the KV cache manager; the multi-agent runtime only
proposes cross-agent pre-prefill tasks and requests offload when background work
needs memory headroom.

\subsection{Cross-Agent Pre-Prefill}

Cross-agent pre-prefill is a multi-agent runtime extension of the
cache-maintenance interface. Unlike stale-coverage prefill, which repairs known
missing or stale coverage for a useful prefix, cross-agent pre-prefill is
speculative and is triggered by pending multi-agent communication. The runtime
uses this state to propose prefill tasks for shared context prefixes before the
downstream agent reaches the foreground path.

A cross-agent pre-prefill proposal contains the target agent, context namespace,
adapter, adapter version, prefix span, currently covered length, estimated token
cost, memory demand, and confidence. The proposal itself is not a separate cache
policy. Instead, it is converted into a predicted-demand signal for the KV cache
manager. For a candidate $c$ with shared prefix $p_c$, target adapter version
$\nu_c=(a_c,v_c)$, missing-token count $\Delta_c$, downstream confidence
$\rho_c$, and $s_c$ pending consumers, the multi-agent runtime reports
\[
    D_{\mathrm{ma}}(c) =
    \rho_c
    \left(1+\beta_s\log(1+s_c)\right)
    \frac{\min(\Delta_c,M_{\max})}{M_{\max}} .
\]
This signal is folded into the prefix heat of $p_c$ and the adapter heat of
$\nu_c$. The final cross-agent pre-prefill priority is computed only after this
multi-agent demand signal is available:
\[
    \Pi_{\mathrm{xpre}}(c)
    =
    D_{\mathrm{ma}}(c)\cdot
    H_{\mathrm{KV}}(e_c)\cdot
    \widehat{C}_{\mathrm{prefill}}(e_c),
    \qquad e_c=(p_c,\nu_c).
\]
Thus, a shared-context prefix is pre-prefilled only when it is both likely to be
reused and likely to be consumed under a relevant adapter version. Full-hit
candidates and candidates with nonpositive predicted demand are discarded; the
remaining candidates are handed to the KV cache manager and admitted only by the
scheduler.

When admitted, a cross-agent pre-prefill task materializes a bounded prefix chunk under the
target adapter version and publishes coverage through the KV cache manager. If a
foreground request reaches the same prefix first, the foreground path performs
ordinary prefill and updates the same metadata. If the target adapter commits a
new version before the prefetched span is consumed, the prefetched coverage is
marked stale and cannot be reused. Thus, cross-agent pre-prefill can move repeated
shared-context computation out of the foreground path, but it never weakens the
versioned-KV correctness rule.

\subsection{Memory-Pressure Offload Requests}

The multi-agent runtime therefore treats memory pressure as a signal to the KV
cache manager, not as a reason to manipulate KV tensors directly. Before it
submits foreground-safe background work, the runtime compares the projected
memory demand of the next task with the scheduler's available headroom. Training
is the most common trigger: a prepared train chunk may need activation and
optimizer memory in addition to resident KV pages. If the projected demand would
consume the foreground reserve, the runtime sends a KV-offload request before
the train task is admitted. The cache manager then applies the priority-based
offload policy in \S\ref{sec:kv-manager}.

The same path handles cache-maintenance work. Before admitting stale-coverage
prefill or cross-agent pre-prefill, the scheduler can ask the KV cache manager
whether the candidate fits the current page budget. Under pressure, the runtime
can shrink a cross-agent pre-prefill chunk, delay a training chunk, restore a
high-priority offloaded span, or request offload of low-priority resident spans.
If the KV manager cannot create enough headroom without touching active adapter
weights or the active request's KV state, the scheduler leaves the background
task pending.

\subsection{Integration with the Core Runtime}

The multi-agent runtime does not introduce a separate model stack. Agent
handles, adapter provenance, and communication state only create additional task
proposals and cache-manager requests. All GPU-bound work still passes through
the cooperative scheduler, all visible adapter updates pass through the adapter
manager, and all reusable KV coverage passes through the version-aware KV cache
manager. A cross-agent pre-prefill task cannot bypass foreground priority, and a
prefetched span is reusable only when its token span, context namespace, adapter,
and adapter version match the eventual foreground request.
\begin{figure*}[t]
    \centering
    \begin{minipage}[t]{0.32\textwidth}
        \centering
        \includegraphics[width=\linewidth]{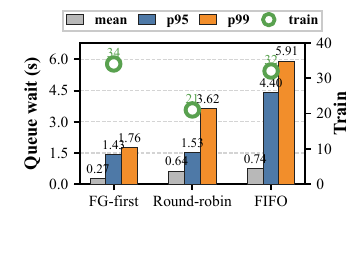}
        \vspace{-0.12in}
        \caption{Scheduling tails.}
        \label{fig:eval-scheduling-baselines}
    \end{minipage}
    \hfill
    \begin{minipage}[t]{0.32\textwidth}
        \centering
        \includegraphics[width=\linewidth]{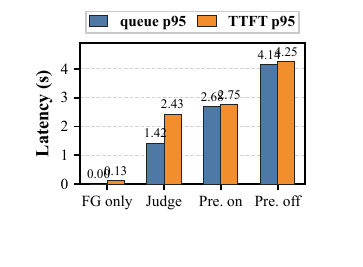}
        \vspace{-0.12in}
        \caption{Training preemption.}
        \label{fig:eval-train-preempt-summary}
    \end{minipage}
    \hfill
    \begin{minipage}[t]{0.32\textwidth}
        \centering
        \includegraphics[width=\linewidth]{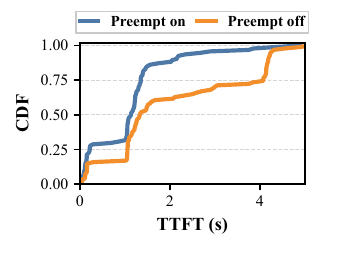}
        \vspace{-0.12in}
        \caption{TTFT under preemption.}
        \label{fig:eval-train-preempt-cdf}
    \end{minipage}
\end{figure*}

In a single-agent deployment, the multi-agent-specific path largely disappears:
the scheduler and KV cache manager still provide version-aware reuse,
stale-coverage prefill, retention, and offload, but no communication structure is
available for cross-agent pre-prefill.

%% file: evaluation.tex
\section{Evaluation} \label{sec:evaluation}

We evaluate the current \name with mechanism-level experiments on a
single consumer GPU. The evaluation asks six questions:

\begin{itemize}[leftmargin=*]
    \item Does the scheduler protect foreground agent turns when judge,
    training, and cache maintenance are also enabled?
    \item How much does interruptible training reduce foreground tail latency?
    \item Is a conventional full-LoRA sleep/train/wake serving path sufficient
    for interactive local learning?
    \item Can stale-coverage prefill reduce the first foreground use after
    adapter updates?
    \item Can cross-agent pre-prefill reduce foreground TTFT when judge
    and training also run?
    \item How does the runtime behave under low VRAM and KV-page pressure?
\end{itemize}

\subsection{Experimental Setup}

Most experiments are 24-task runs on an RTX 3090-class 24GB GPU with CUDA 13.0
and a single model instance. Unless stated otherwise, we use
xLAM-7b-r~\cite{zhang2024xlam,salesforce_xlam_7b_r}, LoRA rank 4, and a judge
budget of 512 output tokens. The main workload is the real tau-bench retail
train split~\cite{yao2024tau} with up to four agent steps per task.
\name is implemented in 24,447 lines of Python code and uses a
FlashInfer-based inference engine~\cite{ye2025flashinfer}. We report
foreground inference latency, foreground queue wait, TTFT, background task
progress, adapter publishes, and peak GPU memory.

We use tau-bench~\cite{yao2024tau} success and reward only to check that runs
complete under the same agent workload. They are not interpreted as
model-quality claims, because the experiments vary runtime policy rather than
agent prompts or model weights.
For latency metrics, we report both the absolute value and the relative change
against the most relevant baseline. A ``$k{\times}$ lower'' tail means that the
baseline tail latency divided by \name's tail latency is $k$.

The scheduler uses a task-cost EMA weight $\alpha_z=0.2$ and a conservative
admission margin $\delta=0.05\,\mathrm{s}$. The cache manager reserves 8
FlashInfer KV pages for foreground work and keeps a
512 MiB safety margin before issuing training offload. Hotness statistics use a
512-access half-life ($\lambda=0.998647$ per access), cap prefix and
missing-token terms at 8192 tokens, and use a shared-context priority default of 2.0 with a
pending-consumer multiplier $\beta_s=0.5$ for cross-agent pre-prefill.

\subsection{Foreground-Aware Scheduling}

\reffig{eval-scheduling-baselines} compares three admission policies on the same
tau-bench workload with foreground inference, judge inference, training, and
stale-coverage prefill enabled. Foreground-first keeps foreground queue wait to
1.433s p95 and 1.758s p99 while still completing 34 training commits and 34
adapter publishes. Round-robin task-type scheduling has a similar p95 of 1.530s
but a larger 3.625s p99 wait and only 21 training commits, because it
continues cycling through background task types even when foreground work
arrives. FIFO is the clearest negative baseline: its p95 queue wait reaches
4.403s and its p99 reaches 5.908s after foreground turns wait behind
already-admitted judge and training work.

In relative terms, foreground-first reduces p95 queue wait by 3.1$\times$
compared with FIFO, a 67.5\% reduction, and reduces p99 queue wait by
3.4$\times$. The comparison with round-robin is more subtle: p95 queue wait is
only 6.4\% lower, but p99 queue wait is 2.1$\times$ lower and the run
finishes 1.6$\times$ more training commits (34 versus 21). This indicates that
round-robin can look acceptable at p95 while still creating larger outliers and
less useful background learning progress.

The queue-attribution logs explain these tails. Foreground-first queue wait is
mostly behind judge work (68.7\%) and training work (18.7\%), with smaller
contributions from stale-coverage prefill and inference. Round-robin spends 68.4\%
behind judge work and 28.8\% behind training. FIFO splits its foreground wait
between training (44.2\%) and judge (44.0\%), with another 9.7\% behind
stale-coverage prefill. These results match the scheduler design: foreground-first
cannot interrupt a GPU kernel that is already running, but it prevents new
background admissions from extending the foreground tail.

\begin{figure*}[t]
    \centering
    \begin{minipage}[t]{0.32\textwidth}
        \centering
        \includegraphics[width=\linewidth]{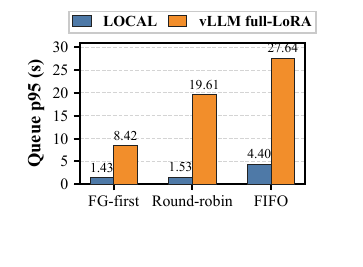}
        \vspace{-0.12in}
        \caption{Sleep/train/wake tails.}
        \label{fig:eval-vllm}
    \end{minipage}
    \hfill
    \begin{minipage}[t]{0.32\textwidth}
        \centering
        \includegraphics[width=\linewidth]{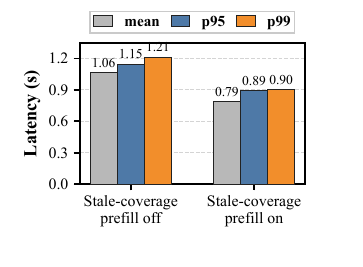}
        \vspace{-0.12in}
        \caption{Stale-coverage first hits.}
        \label{fig:eval-stale-coverage-prefill}
    \end{minipage}
    \hfill
    \begin{minipage}[t]{0.32\textwidth}
        \centering
        \includegraphics[width=\linewidth]{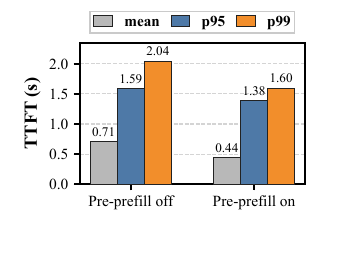}
        \vspace{-0.12in}
        \caption{Cross-agent pre-prefill.}
        \label{fig:eval-shared-prefill}
    \end{minipage}
\end{figure*}

\begin{figure}[t]
    \centering
    \includegraphics[width=0.82\linewidth]{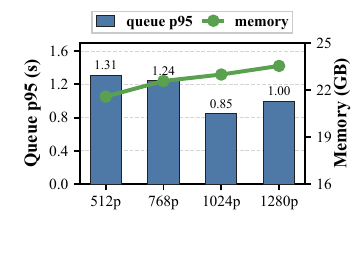}
        \vspace{-0.4in}
    \caption{Memory pressure.}
    \label{fig:eval-memory-pressure}
\end{figure}
\subsection{Interruptible Training}

\reffig{eval-train-preempt-summary} isolates the cost of training preemption on
the tau-bench workload. Foreground-only service has essentially no queueing
tail, with p95 queue wait of 0.0018s. Adding judge inference raises p95 queue
wait to 1.424s. Adding training without preemption raises p95 queue wait to
4.140s and p95 TTFT to 4.247s. With training preemption enabled, p95 queue
wait drops to 2.676s and p95 TTFT drops to 2.749s.

This is the intended trade-off. Interruptible training reduces p95 foreground
queue wait by 1.464s and p95 TTFT by 1.498s, but it also lowers effective
training throughput from 346.8 to 271.8 completed training steps per hour. The
runtime chooses this trade-off because local agents are interactive: background
learning can lose throughput, but it should not hold the only GPU across a
foreground turn boundary. The TTFT CDF in
\reffig{eval-train-preempt-cdf} shows the same tail shift at the request level.
Equivalently, cooperative interruption makes p95 queue wait and p95 TTFT
1.55$\times$ lower than non-preemptible training, while reducing effective
training throughput by 21.6\%. The foreground wait attributed to training drops
from 108.4s to 41.2s across the run, a 2.6$\times$ reduction in accumulated
training-induced blocking.

\subsection{Comparison with Full-LoRA Sleep/Train}

\reffig{eval-vllm} compares \name's custom FlashInfer
runtime against a vLLM full-LoRA sleep/train/wake path on the same workload. In foreground-first
mode, the vLLM path has p95 queue wait of 8.418s, compared with 1.433s in
\name. Under round-robin, vLLM reaches 19.609s p95 queue wait, compared with
1.530s in \name. Under FIFO, vLLM reaches 27.643s p95 queue wait, compared with
4.403s in \name. The median sleep/train/wake window in vLLM is about 12.1s,
including roughly 1.0s to sleep, 9.7s to train, and 1.5s to wake.

Thus, sleep/train/wake makes p95 foreground queue wait 5.9$\times$ higher than
\name under foreground-first admission, 12.8$\times$ higher under round-robin,
and 6.3$\times$ higher under FIFO. The gap comes from the granularity of the
adaptation path: a full sleep/train/wake window is seconds long, while \name
admits judge, training, publish, and cache work as smaller scheduler-visible
units.

This result shows why \name exposes judge, training, publish, and prefill as
scheduler-visible GPU tasks instead of treating local adaptation as a separate
serving mode. Sleep/train/wake reduces peak GPU memory in this run, but the
foreground tail cost is too high for interactive agents. Fine-grained admission
keeps the model available for foreground inference while still letting
background learning progress.

\subsection{Stale-Coverage Prefill}

\reffig{eval-stale-coverage-prefill} evaluates stale-coverage prefill with
foreground-first admission, judge inference, and training. The baseline disables
stale-coverage prefill; the enabled condition
warms up to eight hot prefixes with 128-token chunks after training produces a
new adapter version. Both runs publish more than 30 adapter versions.

Stale-coverage prefill lowers the first foreground prefill under a newly
published adapter. Among exact post-publish first hits, mean prefill falls from
1.064s to 0.789s, p95 falls from 1.146s to 0.894s, and p99 falls from 1.210s to
0.900s. This corresponds to a 25.8\% mean reduction, a 22.0\% p95 reduction,
and a 25.6\% p99 reduction. The measured
first-hit cache coverage p50 rises from 0 to 0.210, showing that the gain comes
from materialized hot-prefix KV rather than a lighter foreground request mix.

\subsection{Cross-Agent Pre-Prefill}

Multi-agent handoffs often reuse a long shared context: one agent observes or
updates the environment, and another consumes the resulting state. \name can
run cross-agent pre-prefill for reusable hot contexts as scheduler-visible
background work before a later foreground turn needs the same prefix.
\reffig{eval-shared-prefill} evaluates this mechanism on the real tau-bench
retail workload with judge inference, training, and adapter publishing enabled.
Both conditions execute the same foreground sequence; the enabled condition adds
43 prefill tasks and 5.184s of prefill GPU time.

Without cross-agent pre-prefill, foreground turns see 0.711s mean TTFT, 1.587s
p95, and 2.042s p99. Enabling cross-agent pre-prefill lowers these metrics to
0.441s, 1.384s, and 1.595s, respectively. This corresponds to a 38.0\% mean
reduction, a 12.8\% p95 reduction, and a 21.9\% p99 reduction. In the full run,
queue attribution shows that prefill rarely blocks foreground work and accounts
for 0.018s of queued foreground time; the dominant blockers remain judge and
training work.

\subsection{Memory Pressure}

\reffig{eval-memory-pressure} sweeps the cache-maintenance parameters that
directly affect admission: KV-page budget, CUDA graph entries, hot-prefix
count, and prefill chunk tokens. On
xLAM-7b-r, all tested configurations
from 512 to 1280 FlashInfer pages complete the workload
without OOM or adapter publish failures. Peak GPU memory ranges from 21.6GB to 23.5GB, and each passing
configuration still completes 31--35 training commits plus 146--283
stale-coverage prefill progress events. The best queue tail among the xLAM runs is
0.850s p95 with 1024 pages, 32 graph entries, four hot prefixes, and 64-token
chunks.

Across the passing xLAM configurations, queue p95 varies by 1.5$\times$ even
though every run has the same model and workload. Increasing the page budget
from 512 to 1024 pages lowers queue p95 from 1.312s to 0.850s, a 1.5$\times$
improvement. Moving to 1280 pages with a larger graph and cache-maintenance
budget raises peak memory to 23.5GB and queue p95 to 0.997s, which is still
below the 512-page and 768-page runs but no longer the best tail. These results
show that memory configuration affects latency through scheduler admission, not
only through whether the run avoids OOM.

Together, these results support the design choice of making memory pressure
visible to the scheduler: foreground latency depends not only on model size,
but also on KV page budget, graph capture budget, hot-prefix admission, and
stale-coverage prefill chunk size.

\subsection{Summary}

The experiments support the two main mechanisms in \name. First, cooperative
scheduling keeps foreground latency bounded under mixed inference, judge,
training, and cache-maintenance work, and interruptible training converts a
large blocking window into a latency-throughput trade-off. Second, the model
runtime can reduce first-use cost after adapter publication, use shared-context
metadata to prefill future consumer contexts during idle windows, and continue
operating under tight KV memory budgets when cache work is admitted
conservatively.

%% file: related_work.tex
\section{Related Work} \label{sec:related}

\textbf{Local inference and LLM serving.}
Local LLM stacks make single-device deployment feasible through portable
runtimes, quantization, and heterogeneous memory use. Existing systems provide
portable local backends, exploit CPU/GPU or flash/DRAM hierarchies, and reduce
model footprint through compression and quantization
~\cite{llamacpp,mlc_llm,executorch,litert_lm,song2023powerinfer,alizadeh2024llmflash,sheng2023flexgen,liu2024mobilellm,frantar2023gptq,lin2024awq,xiao2023smoothquant}.
Serving systems optimize request admission and KV residency under an
inference-only contract: Orca schedules iterations, vLLM manages paged KV,
SGLang exploits structured programs and prefix reuse, and recent systems split
or disaggregate prefill and decode work~\cite{yu2022orca,kwon2023efficient,zheng2024sglangefficientexecutionstructured,agrawal2024sarathi,zhong2024distserve,patel2024splitwise}.
KV-reduction systems such as StreamingLLM, H$_2$O, SnapKV, and KIVI further show
that KV state is a first-order resource~\cite{xiao2024streamingllm,zhang2023h2o,li2024snapkv,liu2024kivi}.
\name shares the same hardware pressure, but the runtime state is mutable:
judge, training, refresh, and foreground inference compete on one GPU, and KV
validity depends on adapter identity and adapter version rather than only on
the prompt and base model.

\textbf{Adapter serving.}
Parameter-efficient methods make local adaptation plausible by updating a small
set of parameters instead of the full model. Adapters, prefix tuning, P-tuning
v2, IA$^3$, LoRA, AdaLoRA, and QLoRA expose different trade-offs between update
cost, memory, and specialization quality~\cite{houlsby2019adapters,li2021prefixtuning,liu2022ptuningv2,liu2022ia3,hu2022lora,zhang2023adalora,dettmers2023qlora}.
Multi-adapter serving systems then batch, place, or share LoRA state for
inference workloads, including Punica, S-LoRA, LoRAX, LRAgent, and
ForkKV~\cite{chen2024punica,sheng2024slora,predibase2023lorax,jeon2026lragent,wang2026forkkv}.
\name does not propose a new adapter method or a new LoRA serving engine.
Instead, it treats adapter identity and adapter version as runtime state that
determines KV validity, cache refresh, and foreground-safe adapter publication.

\textbf{Reinforcement learning and post-training systems.}
RLHF and reinforcement fine-tuning systems organize rollout, reward, reference
scoring, and policy updates. Prior algorithms define policy-gradient,
preference-learning, and reasoning-RL objectives
~\cite{schulman2017ppo,ouyang2022instructgpt,bai2022constitutional,rafailov2023dpo,deepseekai2025deepseekr1incentivizingreasoningcapability}.
Distributed RL and LLM post-training frameworks then scale these stages across
separate rollout, reward, reference, and trainer workers, using systems such as
Ray and recent RLHF/RFT frameworks to optimize cluster throughput
~\cite{moritz2018ray,sheng2024hybridflow,hu2024openrlhf,mei2025real,fu2025areal,slime_github,pan2025trinityrft}.
OpenClaw-RL is closer in spirit to agent-facing reinforcement learning because
it studies learning from interaction feedback in an agent setting
~\cite{wang2026openclawrl}. \name is complementary: it targets the
single-device runtime case where rollout, judging, cache maintenance, and
adapter updates are not placed on separate workers, so background learning must
become scheduler-visible, interruptible, and memory-aware.

\textbf{LLM agents and multi-agent runtimes.}
Agent systems study tool use, environment interaction, delayed feedback, and
coordination. ReAct, Toolformer, MRKL, WebGPT, SayCan, ToolLLM, and AgentBench
connect LLMs to tools, web or embodied actions, and evaluation
tasks~\cite{yao2023react,schick2023toolformer,karpas2022mrkl,nakano2021webgpt,ahn2022saycan,qin2023toollm,liu2023agentbench}.
Reflexion, Voyager, and OpenClaw-RL use later observations or feedback to
improve behavior~\cite{shinn2023reflexion,wang2023voyager,wang2026openclawrl},
and CAMEL, AutoGen, MetaGPT, and generative agents organize applications around
roles, messages, shared context, and collaboration~\cite{li2023camel,wu2023autogen,hong2024metagpt,park2023generativeagents}.
These frameworks usually treat orchestration metadata as application state.
\name treats the same metadata as runtime state: it drives feedback
attribution, adapter binding, cross-agent pre-prefill, and safe KV reuse on a
single constrained GPU.

%% file: conclusion.tex
\section{Conclusion} \label{sec:conclusion}

On-device interactive RL breaks the inference-only assumption behind many local
LLM runtimes, preventing agents from learning contiguously from their
interactions. Foreground inference, delayed reward recovery, adapter training,
and KV-cache lifecycle management all compete for one GPU and one memory
budget. \name addresses this mismatch through three runtime components:
a cooperative scheduler, a version-aware KV-cache manager, and a multi-agent
model runtime that carries agent, adapter, feedback, and KV provenance through
serving and learning.

The scheduler centralizes GPU admission, protects foreground inference,
supports interruptible training, and admits judge, training, and cache
refresh work only when the runtime can account for their cost. The KV-cache
manager preserves versioned cache correctness and manages retention, offload,
and refresh under memory pressure. The model runtime preserves agent identity,
communication context, adapter attribution, and adapter-scoped KV state across
interaction, judging, replay, and adaptation.

Our prototype evaluation shows that these mechanisms keep foreground queue
tails bounded under mixed agent workloads, reduce the tail cost of training via
interruption, improve first-consumer latency for shared-context handoffs during
idle windows, and continue to make background progress under tight KV memory
budgets. The remaining work is to broaden the evidence to longer multi-agent
sessions, larger adapter populations, and policy-level learning quality.

%% file: reference.bib
@misc{mlc_llm,
  author = {{MLC Team}},
  title = {{MLC-LLM}: Universal {LLM} Deployment Engine with {ML} Compilation},
  year = {2023},
  howpublished = {\url{https://github.com/mlc-ai/mlc-llm}},
  note = {Accessed: 2026-04-29}
}

@misc{executorch,
  author = {{PyTorch Team}},
  title = {{ExecuTorch}: End-to-End Solution for On-Device Inference},
  year = {2024},
  howpublished = {\url{https://pytorch.org/executorch/}},
  note = {Accessed: 2026-06-07}
}

@misc{litert_lm,
  author = {{Google AI Edge}},
  title = {{LiteRT-LM}: Run {LLMs} on Device},
  year = {2026},
  howpublished = {\url{https://ai.google.dev/edge/litert/litert-lm/overview}},
  note = {Accessed: 2026-06-11}
}

@inproceedings{song2023powerinfer,
  title = {{PowerInfer}: Fast Large Language Model Serving with a Consumer-grade {GPU}},
  author = {Song, Yixin and Mi, Zeyu and Xie, Haotong and Chen, Haibo},
  booktitle = {Proceedings of the ACM SIGOPS 30th Symposium on Operating Systems Principles},
  year = {2024},
  url = {https://arxiv.org/abs/2312.12456}
}

@inproceedings{alizadeh2024llmflash,
  title = {{LLM} in a Flash: Efficient Large Language Model Inference with Limited Memory},
  author = {Alizadeh, Keivan and Mirzadeh, Iman and Belenko, Dmitry and Khatamifard, Karen and Cho, Minsik and Del Mundo, Carlo C. and Rastegari, Mohammad and Farajtabar, Mehrdad},
  booktitle = {Proceedings of the 62nd Annual Meeting of the Association for Computational Linguistics},
  year = {2024}
}

@inproceedings{ye2025flashinfer,
  title = {{FlashInfer}: Efficient and Customizable Attention Engine for {LLM} Inference Serving},
  author = {Ye, Zihao and Chen, Lequn and Lai, Ruihang and Lin, Wuwei and Zhang, Yineng and Wang, Stephanie and Chen, Tianqi and Kasikci, Baris and Grover, Vinod and Krishnamurthy, Arvind and Ceze, Luis},
  booktitle = {Proceedings of Machine Learning and Systems},
  year = {2025},
  url = {https://arxiv.org/abs/2501.01005}
}

@article{zhang2024xlam,
  title = {{xLAM}: A Family of Large Action Models to Empower {AI} Agent Systems},
  author = {Zhang, Jianguo and Lan, Tian and Zhu, Ming and Liu, Zuxin and Hoang, Thai and Kokane, Shirley and Yao, Weiran and Tan, Juntao and Prabhakar, Akshara and Chen, Haolin and Liu, Zhiwei and Feng, Yihao and Awalgaonkar, Tulika and Murthy, Rithesh and Hu, Eric and Chen, Zeyuan and Xu, Ran and Niebles, Juan Carlos and Heinecke, Shelby and Wang, Huan and Savarese, Silvio and Xiong, Caiming},
  journal = {arXiv preprint arXiv:2409.03215},
  year = {2024},
  url = {https://arxiv.org/abs/2409.03215}
}

@misc{salesforce_xlam_7b_r,
  author = {{Salesforce AI Research}},
  title = {{xLAM-7b-r}},
  year = {2024},
  howpublished = {\url{https://huggingface.co/Salesforce/xLAM-7b-r}},
  note = {Model card. Accessed: 2026-06-11}
}

@article{yao2024tau,
  title = {{$\tau$-bench}: A Benchmark for Tool-Agent-User Interaction in Real-World Domains},
  author = {Yao, Shunyu and Shinn, Noah and Razavi, Pedram and Narasimhan, Karthik},
  journal = {arXiv preprint arXiv:2406.12045},
  year = {2024},
  url = {https://arxiv.org/abs/2406.12045}
}

@inproceedings{liu2024mobilellm,
  title = {{MobileLLM}: Optimizing Sub-billion Parameter Language Models for On-Device Use Cases},
  author = {Liu, Zechun and Zhao, Changsheng and Iandola, Forrest and Lai, Chen and Tian, Yuandong and Fedorov, Igor and Xiong, Yunyang and Chang, Ernie and Shi, Yangyang and Krishnamoorthi, Raghuraman and Lai, Liangzhen and Chandra, Vikas},
  booktitle = {International Conference on Machine Learning},
  year = {2024}
}

@inproceedings{hu2022lora,
  title = {{LoRA}: Low-Rank Adaptation of Large Language Models},
  author = {Hu, Edward J. and Shen, Yelong and Wallis, Phillip and Allen-Zhu, Zeyuan and Li, Yuanzhi and Wang, Shean and Wang, Lu and Chen, Weizhu},
  booktitle = {International Conference on Learning Representations},
  year = {2022},
  url = {https://openreview.net/forum?id=nZeVKeeFYf9}
}

@inproceedings{kwon2023efficient,
  title = {Efficient Memory Management for Large Language Model Serving with {PagedAttention}},
  author = {Kwon, Woosuk and Li, Zhuohan and Zhuang, Siyuan and Sheng, Ying and Zheng, Lianmin and Yu, Cody Hao and Gonzalez, Joseph E. and Zhang, Hao and Stoica, Ion},
  booktitle = {Proceedings of the ACM SIGOPS 29th Symposium on Operating Systems Principles},
  pages = {611--626},
  year = {2023},
  doi = {10.1145/3600006.3613165},
  url = {https://arxiv.org/abs/2309.06180}
}

@inproceedings{zheng2024sglangefficientexecutionstructured,
  title = {{SGLang}: Efficient Execution of Structured Language Model Programs},
  author = {Zheng, Lianmin and Yin, Liangsheng and Xie, Zhiqiang and Sun, Chuyue and Huang, Jeff and Yu, Cody Hao and Cao, Shiyi and Kozyrakis, Christos and Stoica, Ion and Gonzalez, Joseph E. and Barrett, Clark and Sheng, Ying},
  booktitle = {Advances in Neural Information Processing Systems},
  year = {2024},
  url = {https://arxiv.org/abs/2312.07104}
}

@article{deepseekai2025deepseekr1incentivizingreasoningcapability,
  title = {{DeepSeek-R1}: Incentivizing Reasoning Capability in {LLMs} via Reinforcement Learning},
  author = {{DeepSeek-AI}},
  journal = {arXiv preprint arXiv:2501.12948},
  year = {2025},
  url = {https://arxiv.org/abs/2501.12948}
}

@article{wang2026openclawrl,
  title = {{OpenClaw-RL}: Train Any Agent Simply by Talking},
  author = {Wang, Yinjie and Chen, Xuyang and Jin, Xiaolong and Wang, Mengdi and Yang, Ling},
  journal = {arXiv preprint arXiv:2603.10165},
  year = {2026},
  url = {https://arxiv.org/abs/2603.10165}
}

@inproceedings{gao2025mobizo,
  title = {{MobiZO}: Enabling Efficient {LLM} Fine-Tuning at the Edge via Inference Engines},
  author = {Gao, Lei and Ziashahabi, Amir and Niu, Yue and Avestimehr, Salman and Annavaram, Murali},
  booktitle = {Proceedings of the 2025 Conference on Empirical Methods in Natural Language Processing},
  pages = {20206--20223},
  year = {2025},
  address = {Suzhou, China},
  publisher = {Association for Computational Linguistics},
  doi = {10.18653/v1/2025.emnlp-main.1022},
  url = {https://aclanthology.org/2025.emnlp-main.1022/}
}

@inproceedings{jeon2026lragent,
  title = {{LRAgent}: Efficient {KV} Cache Sharing for Multi-{LoRA} {LLM} Agents},
  author = {Jeon, Hyesung and Ha, Hyeongju and Kim, Jae-Joon},
  booktitle = {International Conference on Machine Learning},
  year = {2026},
  url = {https://arxiv.org/abs/2602.01053}
}

@article{wang2026forkkv,
  title = {{ForkKV}: Scaling Multi-{LoRA} Agent Serving via Copy-on-Write Disaggregated {KV} Cache},
  author = {Wang, Shao and Ren, Rui and Gui, Lin},
  journal = {arXiv preprint arXiv:2604.06370},
  year = {2026},
  url = {https://arxiv.org/abs/2604.06370}
}

@inproceedings{sheng2024hybridflow,
  title = {{HybridFlow}: A Flexible and Efficient {RLHF} Framework},
  author = {Sheng, Guangming and Zhang, Chi and Ye, Zilingfeng and Wu, Xibin and Zhang, Wang and Zhang, Ru and Peng, Yanghua and Lin, Haibin and Wu, Chuan},
  booktitle = {Proceedings of the Twentieth European Conference on Computer Systems},
  year = {2025},
  doi = {10.1145/3689031.3696075},
  url = {https://doi.org/10.1145/3689031.3696075}
}

@inproceedings{mei2025real,
  title = {{ReaL}: Efficient {RLHF} Training of Large Language Models with Parameter Reallocation},
  author = {Mei, Zhiyu and Fu, Wei and Li, Kaiwei and Wang, Guangju and Zhang, Huanchen and Wu, Yi},
  booktitle = {Proceedings of Machine Learning and Systems},
  volume = {7},
  year = {2025},
  url = {https://proceedings.mlsys.org/paper_files/paper/2025/hash/3b3889d313ba9476c12c2d77ea66b24f-Abstract-Conference.html}
}

@article{hu2024openrlhf,
  title = {{OpenRLHF}: An Easy-to-use, Scalable and High-performance {RLHF} Framework},
  author = {Hu, Jian and Wu, Xibin and Zhu, Zilin and Xianyu and Wang, Weixun and Zhang, Dehao and Cao, Yu},
  journal = {arXiv preprint arXiv:2405.11143},
  year = {2024}
}

@misc{slime_github,
  author = {{THUDM}},
  title = {{slime}: An {LLM} Post-Training Framework for {RL} Scaling},
  year = {2025},
  howpublished = {\url{https://github.com/THUDM/slime}},
  note = {GitHub repository. Accessed: 2026-06-08}
}

@article{fu2025areal,
  title = {{AReaL}: A Large-Scale Asynchronous Reinforcement Learning System for Language Reasoning},
  author = {Fu, Wei and Gao, Jiaxuan and Shen, Xujie and Zhu, Chen and Mei, Zhiyu and He, Chuyi and Xu, Shusheng and Wei, Guo and Mei, Jun and Wang, Jiashu and Yang, Tongkai and Yuan, Binhang and Wu, Yi},
  journal = {arXiv preprint arXiv:2505.24298},
  year = {2025}
}

@article{pan2025trinityrft,
  title = {{Trinity-RFT}: A General-Purpose and Unified Framework for Reinforcement Fine-Tuning of Large Language Models},
  author = {Pan, Xuchen and Chen, Yanxi and Chen, Yushuo and Sun, Yuchang and Chen, Daoyuan and Zhang, Wenhao and Xie, Yuexiang and Huang, Yilun and Zhang, Yilei and Gao, Dawei and Shi, Weijie and Li, Yaliang and Ding, Bolin and Zhou, Jingren},
  journal = {arXiv preprint arXiv:2505.17826},
  year = {2025},
  url = {https://arxiv.org/abs/2505.17826}
}

@inproceedings{yao2023react,
  title = {{ReAct}: Synergizing Reasoning and Acting in Language Models},
  author = {Yao, Shunyu and Zhao, Jeffrey and Yu, Dian and Du, Nan and Shafran, Izhak and Narasimhan, Karthik and Cao, Yuan},
  booktitle = {International Conference on Learning Representations},
  year = {2023},
  url = {https://arxiv.org/abs/2210.03629}
}

@inproceedings{schick2023toolformer,
  title = {{Toolformer}: Language Models Can Teach Themselves to Use Tools},
  author = {Schick, Timo and Dwivedi-Yu, Jane and Dess{\`i}, Roberto and Raileanu, Roberta and Lomeli, Maria and Hambro, Eric and Zettlemoyer, Luke and Cancedda, Nicola and Scialom, Thomas},
  booktitle = {Advances in Neural Information Processing Systems},
  year = {2023},
  url = {https://arxiv.org/abs/2302.04761}
}

@inproceedings{shinn2023reflexion,
  title = {{Reflexion}: Language Agents with Verbal Reinforcement Learning},
  author = {Shinn, Noah and Cassano, Federico and Gopinath, Ashwin and Narasimhan, Karthik and Yao, Shunyu},
  booktitle = {Advances in Neural Information Processing Systems},
  year = {2023},
  url = {https://arxiv.org/abs/2303.11366}
}

@article{wang2023voyager,
  title = {{Voyager}: An Open-Ended Embodied Agent with Large Language Models},
  author = {Wang, Guanzhi and Xie, Yuqi and Jiang, Yunfan and Mandlekar, Ajay and Xiao, Chaowei and Zhu, Yuke and Fan, Linxi and Anandkumar, Anima},
  journal = {arXiv preprint arXiv:2305.16291},
  year = {2023},
  url = {https://arxiv.org/abs/2305.16291}
}

@article{li2023camel,
  title = {{CAMEL}: Communicative Agents for ``Mind'' Exploration of Large Language Model Society},
  author = {Li, Guohao and Hammoud, Hasan Abed Al Kader and Itani, Hani and Khizbullin, Dmitrii and Ghanem, Bernard},
  journal = {arXiv preprint arXiv:2303.17760},
  year = {2023}
}

@inproceedings{wu2023autogen,
  title = {{AutoGen}: Enabling Next-Gen {LLM} Applications via Multi-Agent Conversation Framework},
  author = {Wu, Qingyun and Bansal, Gagan and Zhang, Jieyu and Wu, Yiran and Li, Shaokun and Zhu, Erkang and Jiang, Beibin and Zhang, Li and Zhang, Xiaoyun and Awadallah, Ahmed Hassan and White, Ryen W. and Burger, Doug and Wang, Chi},
  booktitle = {Conference on Language Modeling},
  year = {2024},
  url = {https://arxiv.org/abs/2308.08155}
}

@inproceedings{hong2024metagpt,
  title = {{MetaGPT}: Meta Programming for A Multi-Agent Collaborative Framework},
  author = {Hong, Sirui and Zhuge, Mingchen and Chen, Jiaqi and Zheng, Xiawu and Cheng, Yuheng and Zhang, Ceyao and Wang, Jinlin and Wang, Zili and Yau, Steven Ka Shing and Lin, Zijuan and Zhou, Liyang and Ran, Chenyu and Xiao, Lingfeng and Wu, Chenglin and Schmidhuber, J{\"u}rgen},
  booktitle = {International Conference on Learning Representations},
  year = {2024}
}

@misc{llamacpp,
  author = {Gerganov, Georgi},
  title = {{llama.cpp}: Port of {LLaMA} model in {C/C++}},
  year = {2023},
  howpublished = {\url{https://github.com/ggml-org/llama.cpp}},
  note = {GitHub repository. Accessed: 2026-06-08}
}

@inproceedings{sheng2023flexgen,
  title = {{FlexGen}: High-Throughput Generative Inference of Large Language Models with a Single {GPU}},
  author = {Sheng, Ying and Zheng, Lianmin and Yuan, Binhang and Li, Zhuohan and Ryabinin, Max and Fu, Daniel Y. and Xie, Zhiqiang and Chen, Beidi and Barrett, Clark and Gonzalez, Joseph E. and Liang, Percy and R{\'e}, Christopher and Stoica, Ion and Zhang, Ce},
  booktitle = {International Conference on Machine Learning},
  year = {2023},
  url = {https://arxiv.org/abs/2303.06865}
}

@inproceedings{frantar2023gptq,
  title = {{GPTQ}: Accurate Post-Training Quantization for Generative Pre-trained Transformers},
  author = {Frantar, Elias and Ashkboos, Saleh and Hoefler, Torsten and Alistarh, Dan},
  booktitle = {International Conference on Learning Representations},
  year = {2023}
}

@inproceedings{lin2024awq,
  title = {{AWQ}: Activation-aware Weight Quantization for {LLM} Compression and Acceleration},
  author = {Lin, Ji and Tang, Jiaming and Tang, Haotian and Yang, Shang and Chen, Wei-Ming and Wang, Wei-Chen and Xiao, Guangxuan and Dang, Xingyu and Gan, Chuang and Han, Song},
  booktitle = {Proceedings of Machine Learning and Systems},
  volume = {6},
  year = {2024},
  url = {https://proceedings.mlsys.org/paper_files/paper/2024/hash/42a452cbafa9dd64e9ba4aa95cc1ef21-Abstract-Conference.html}
}

@inproceedings{xiao2023smoothquant,
  title = {{SmoothQuant}: Accurate and Efficient Post-Training Quantization for Large Language Models},
  author = {Xiao, Guangxuan and Lin, Ji and Seznec, Mickael and Wu, Hao and Demouth, Julien and Han, Song},
  booktitle = {International Conference on Machine Learning},
  year = {2023},
  url = {https://arxiv.org/abs/2211.10438}
}

@inproceedings{yu2022orca,
  title = {{Orca}: A Distributed Serving System for Transformer-Based Generative Models},
  author = {Yu, Gyeong-In and Jeong, Joo Seong and Kim, Geon-Woo and Kim, Soojeong and Chun, Byung-Gon},
  booktitle = {16th USENIX Symposium on Operating Systems Design and Implementation},
  pages = {521--538},
  year = {2022},
  url = {https://www.usenix.org/conference/osdi22/presentation/yu}
}

@inproceedings{agrawal2024sarathi,
  title = {Taming Throughput-Latency Tradeoff in {LLM} Inference with {Sarathi-Serve}},
  author = {Agrawal, Amey and Kedia, Nitin and Panwar, Ashish and Mohan, Jayashree and Kwatra, Nipun and Gulavani, Bhargav S. and Tumanov, Alexey and Ramjee, Ramachandran},
  booktitle = {18th USENIX Symposium on Operating Systems Design and Implementation},
  pages = {117--134},
  year = {2024},
  url = {https://arxiv.org/abs/2403.02310}
}

@inproceedings{zhong2024distserve,
author = {Yinmin Zhong and Shengyu Liu and Junda Chen and Jianbo Hu and Yibo Zhu and Xuanzhe Liu and Xin Jin and Hao Zhang},
title = {{DistServe}: Disaggregating Prefill and Decoding for Goodput-optimized Large Language Model Serving},
booktitle = {18th USENIX Symposium on Operating Systems Design and Implementation (OSDI 24)},
year = {2024},
isbn = {978-1-939133-40-3},
address = {Santa Clara, CA},
pages = {193--210},
url = {https://www.usenix.org/conference/osdi24/presentation/zhong-yinmin},
publisher = {USENIX Association},
month = jul
}

@inproceedings{patel2024splitwise,
author = {Patel, Pratyush and Choukse, Esha and Zhang, Chaojie and Shah, Aashaka and Goiri, \'{I}\~{n}igo and Maleki, Saeed and Bianchini, Ricardo},
title = {Splitwise: Efficient Generative LLM Inference Using Phase Splitting},
year = {2025},
isbn = {9798350326581},
publisher = {IEEE Press},
url = {https://doi.org/10.1109/ISCA59077.2024.00019},
doi = {10.1109/ISCA59077.2024.00019},
booktitle = {Proceedings of the 51st Annual International Symposium on Computer Architecture},
pages = {118–132},
numpages = {15},
location = {Buenos Aires, Argentina},
series = {ISCA '24}
}

@inproceedings{xiao2024streamingllm,
 author = {Xiao, Guangxuan and Tian, Yuandong and Chen, Beidi and Han, Song and Lewis, Mike },
 booktitle = {International Conference on Learning Representations},
 editor = {B. Kim and Y. Yue and S. Chaudhuri and K. Fragkiadaki and M. Khan and Y. Sun},
 pages = {21875--21895},
 title = {Efficient Streaming Language Models with Attention Sinks},
 url = {https://proceedings.iclr.cc/paper_files/paper/2024/file/5e5fd18f863cbe6d8ae392a93fd271c9-Paper-Conference.pdf},
 volume = {2024},
 year = {2024}
}

@inproceedings{zhang2023h2o,
 author = {Zhang, Zhenyu and Sheng, Ying and Zhou, Tianyi and Chen, Tianlong and Zheng, Lianmin and Cai, Ruisi and Song, Zhao and Tian, Yuandong and R\'{e}, Christopher and Barrett, Clark and Wang, Zhangyang "Atlas" and Chen, Beidi},
 booktitle = {Advances in Neural Information Processing Systems},
 editor = {A. Oh and T. Naumann and A. Globerson and K. Saenko and M. Hardt and S. Levine},
 pages = {34661--34710},
 publisher = {Curran Associates, Inc.},
 title = {H2O: Heavy-Hitter Oracle for Efficient Generative Inference of Large Language Models},
 url = {https://proceedings.neurips.cc/paper_files/paper/2023/file/6ceefa7b15572587b78ecfcebb2827f8-Paper-Conference.pdf},
 volume = {36},
 year = {2023}
}

@inproceedings{li2024snapkv,
  title = {{SnapKV}: {LLM} Knows What You are Looking for Before Generation},
  author = {Li, Yuhong and Huang, Yingbing and Yang, Bowen and Venkitesh, Bharat and Locatelli, Andrea and Ye, Hanchen and Cai, Tianle and Lewis, Patrick and Chen, Deming},
  booktitle = {Advances in Neural Information Processing Systems},
  year = {2024},
  url = {https://arxiv.org/abs/2404.14469}
}

@inproceedings{liu2024kivi,
  title = 	 {{KIVI}: A Tuning-Free Asymmetric 2bit Quantization for {KV} Cache},
  author =       {Liu, Zirui and Yuan, Jiayi and Jin, Hongye and Zhong, Shaochen and Xu, Zhaozhuo and Braverman, Vladimir and Chen, Beidi and Hu, Xia},
  booktitle = 	 {Proceedings of the 41st International Conference on Machine Learning},
  pages = 	 {32332--32344},
  year = 	 {2024},
  editor = 	 {Salakhutdinov, Ruslan and Kolter, Zico and Heller, Katherine and Weller, Adrian and Oliver, Nuria and Scarlett, Jonathan and Berkenkamp, Felix},
  volume = 	 {235},
  series = 	 {Proceedings of Machine Learning Research},
  month = 	 {21--27 Jul},
  publisher =    {PMLR},
  url = 	 {https://proceedings.mlr.press/v235/liu24bz.html},
}

@inproceedings{houlsby2019adapters,
  title = {Parameter-Efficient Transfer Learning for {NLP}},
  author = {Houlsby, Neil and Giurgiu, Andrei and Jastrzebski, Stanislaw and Morrone, Bruna and de Laroussilhe, Quentin and Gesmundo, Andrea and Attariyan, Mona and Gelly, Sylvain},
  booktitle = {International Conference on Machine Learning},
  year = {2019},
  url = {https://arxiv.org/abs/1902.00751}
}

@inproceedings{li2021prefixtuning,
  title = {Prefix-Tuning: Optimizing Continuous Prompts for Generation},
  author = {Li, Xiang Lisa and Liang, Percy},
  booktitle = {Proceedings of the 59th Annual Meeting of the Association for Computational Linguistics},
  year = {2021},
  url = {https://arxiv.org/abs/2101.00190}
}

@inproceedings{liu2022ptuningv2,
  title = {{P-Tuning} v2: Prompt Tuning Can Be Comparable to Fine-tuning Universally Across Scales and Tasks},
  author = {Liu, Xiao and Ji, Kaixuan and Fu, Yicheng and Tam, Weng Lam and Du, Zhengxiao and Yang, Zhilin and Tang, Jie},
  booktitle = {Proceedings of the 60th Annual Meeting of the Association for Computational Linguistics},
  year = {2022},
  url = {https://arxiv.org/abs/2110.07602}
}

@inproceedings{liu2022ia3,
  title = {Few-Shot Parameter-Efficient Fine-Tuning is Better and Cheaper than In-Context Learning},
  author = {Liu, Haokun and Tam, Derek and Mohammed, Muqeeth and Mohta, Jay and Huang, Tenghao and Bansal, Mohit and Raffel, Colin A.},
  booktitle = {Advances in Neural Information Processing Systems},
  year = {2022},
  url = {https://arxiv.org/abs/2205.05638}
}

@inproceedings{zhang2023adalora,
title={Adaptive Budget Allocation for Parameter-Efficient Fine-Tuning },
author={Qingru Zhang and Minshuo Chen and Alexander Bukharin and Pengcheng He and Yu Cheng and Weizhu Chen and Tuo Zhao},
booktitle={The Eleventh International Conference on Learning Representations },
year={2023},
url={https://openreview.net/forum?id=lq62uWRJjiY}
}

@inproceedings{dettmers2023qlora,
  title = {{QLoRA}: Efficient Finetuning of Quantized {LLMs}},
  author = {Dettmers, Tim and Pagnoni, Artidoro and Holtzman, Ari and Zettlemoyer, Luke},
  booktitle = {Advances in Neural Information Processing Systems},
  year = {2023},
  url = {https://arxiv.org/abs/2305.14314}
}

@inproceedings{chen2024punica,
  title = {{Punica}: Multi-Tenant {LoRA} Serving},
  author = {Chen, Lequn and Ye, Zihao and Wu, Yizhuo and Zhuo, Danyang and Ceze, Luis and Krishnamurthy, Arvind},
  booktitle = {Proceedings of Machine Learning and Systems},
  volume = {6},
  year = {2024},
  url = {https://arxiv.org/abs/2310.18547}
}

@inproceedings{sheng2024slora,
 author = {Sheng, Ying and Cao, Shiyi and Li, Dacheng and Hooper, Coleman and Lee, Nicholas and Yang, Shuo and Chou, Christopher and Zhu, Banghua and Zheng, Lianmin and Keutzer, Kurt and Gonzalez, Joseph E. and Stoica, Ion},
 booktitle = {Proceedings of Machine Learning and Systems},
 editor = {P. Gibbons and G. Pekhimenko and C. De Sa},
 pages = {296--311},
 title = {SLoRA: Scalable Serving of Thousands of LoRA Adapters},
 url = {https://proceedings.mlsys.org/paper_files/paper/2024/file/906419cd502575b617cc489a1a696a67-Paper-Conference.pdf},
 volume = {6},
 year = {2024}
}

@misc{predibase2023lorax,
  author = {{Predibase}},
  title = {{LoRAX}: Multi-{LoRA} Inference Server},
  year = {2023},
  howpublished = {\url{https://github.com/predibase/lorax}},
  note = {GitHub repository. Accessed: 2026-06-08}
}

@article{schulman2017ppo,
  title = {Proximal Policy Optimization Algorithms},
  author = {Schulman, John and Wolski, Filip and Dhariwal, Prafulla and Radford, Alec and Klimov, Oleg},
  journal = {arXiv preprint arXiv:1707.06347},
  year = {2017},
  url = {https://arxiv.org/abs/1707.06347}
}

@inproceedings{ouyang2022instructgpt,
 author = {Ouyang, Long and Wu, Jeffrey and Jiang, Xu and Almeida, Diogo and Wainwright, Carroll and Mishkin, Pamela and Zhang, Chong and Agarwal, Sandhini and Slama, Katarina and Ray, Alex and Schulman, John and Hilton, Jacob and Kelton, Fraser and Miller, Luke and Simens, Maddie and Askell, Amanda and Welinder, Peter and Christiano, Paul F and Leike, Jan and Lowe, Ryan},
 booktitle = {Advances in Neural Information Processing Systems},
 editor = {S. Koyejo and S. Mohamed and A. Agarwal and D. Belgrave and K. Cho and A. Oh},
 pages = {27730--27744},
 publisher = {Curran Associates, Inc.},
 title = {Training language models to follow instructions with human feedback},
 url = {https://proceedings.neurips.cc/paper_files/paper/2022/file/b1efde53be364a73914f58805a001731-Paper-Conference.pdf},
 volume = {35},
 year = {2022}
}

@article{bai2022constitutional,
  title = {Constitutional {AI}: Harmlessness from {AI} Feedback},
  author = {Bai, Yuntao and Kadavath, Saurav and Kundu, Sandipan and Askell, Amanda and Kernion, Jackson and Jones, Andy and Chen, Anna and Goldie, Anna and Mirhoseini, Azalia and McKinnon, Cameron and Chen, Carol and Olsson, Catherine and Olah, Christopher and Hernandez, Danny and Drain, Dawn and Ganguli, Deep and Li, Dustin and Tran-Johnson, Eli and Perez, Ethan and Kerr, Jamie and Mueller, Jared and Ladish, Jeffrey and Landau, Joshua and Ndousse, Kamal and Lukosiute, Kamile and Lovitt, Liane and Sellitto, Michael and Elhage, Nelson and Schiefer, Nicholas and Mercado, Noem{\'i} and DasSarma, Nova and Lasenby, Robert and Larson, Robin and Ringer, Sam and Johnston, Scott and Kravec, Shauna and Showk, Sheer El and Fort, Stanislav and Lanham, Tamera and Telleen-Lawton, Timothy and Conerly, Tom and Henighan, Tom and Hume, Tristan and Bowman, Samuel R. and Hatfield-Dodds, Zac and Mann, Ben and Amodei, Dario and Joseph, Nicholas and McCandlish, Sam and Brown, Tom and Kaplan, Jared},
  journal = {arXiv preprint arXiv:2212.08073},
  year = {2022},
  url = {https://arxiv.org/abs/2212.08073}
}

@inproceedings{rafailov2023dpo,
 author = {Rafailov, Rafael and Sharma, Archit and Mitchell, Eric and Manning, Christopher D and Ermon, Stefano and Finn, Chelsea},
 booktitle = {Advances in Neural Information Processing Systems},
 editor = {A. Oh and T. Naumann and A. Globerson and K. Saenko and M. Hardt and S. Levine},
 pages = {53728--53741},
 publisher = {Curran Associates, Inc.},
 title = {Direct Preference Optimization: Your Language Model is Secretly a Reward Model},
 url = {https://proceedings.neurips.cc/paper_files/paper/2023/file/a85b405ed65c6477a4fe8302b5e06ce7-Paper-Conference.pdf},
 volume = {36},
 year = {2023}
}

@inproceedings{moritz2018ray,
  title = {{Ray}: A Distributed Framework for Emerging {AI} Applications},
  author = {Moritz, Philipp and Nishihara, Robert and Wang, Stephanie and Tumanov, Alexey and Liaw, Richard and Liang, Eric and Elibol, Melih and Yang, Zongheng and Paul, William and Jordan, Michael I. and Stoica, Ion},
  booktitle = {13th USENIX Symposium on Operating Systems Design and Implementation},
  pages = {561--577},
  year = {2018},
  url = {https://www.usenix.org/conference/osdi18/presentation/moritz}
}

@article{karpas2022mrkl,
  title = {{MRKL} Systems: A Modular, Neuro-Symbolic Architecture That Combines Large Language Models, External Knowledge Sources and Discrete Reasoning},
  author = {Karpas, Eyal and Abend, Omri and Belinkov, Yonatan and Lenz, Barak and Lieber, Opher and Ratner, Nir and Shoham, Yoav and Bata, Yoav and Levine, Yoav and Leyton-Brown, Kevin and Muhlgay, Dor and Roit, Paul and Shashua, Amnon},
  journal = {arXiv preprint arXiv:2205.00445},
  year = {2022},
  url = {https://arxiv.org/abs/2205.00445}
}

@article{nakano2021webgpt,
  title = {{WebGPT}: Browser-assisted Question-Answering with Human Feedback},
  author = {Nakano, Reiichiro and Hilton, Jacob and Balaji, Suchir and Wu, Jeff and Ouyang, Long and Kim, Christina and Hesse, Christopher and Jain, Shantanu and Kosaraju, Vineet and Saunders, William and Jiang, Xu and Cobbe, Karl and Eloundou, Tyna and Krueger, Gretchen and Button, Kevin and Knight, Matthew and Chess, Benjamin and Schulman, John},
  journal = {arXiv preprint arXiv:2112.09332},
  year = {2021},
  url = {https://arxiv.org/abs/2112.09332}
}

@article{ahn2022saycan,
  title = {Do As I Can, Not As I Say: Grounding Language in Robotic Affordances},
  author = {Ahn, Michael and Brohan, Anthony and Brown, Noah and Chebotar, Yevgen and Cortes, Omar and David, Byron and Finn, Chelsea and Fu, Chuyuan and Gopalakrishnan, Keerthana and Hausman, Karol and Herzog, Alexander and Ho, Daniel and Hsu, Jasmine and Ibarz, Julian and Ichter, Brian and Irpan, Alex and Jang, Eric and Ruano, Rosario Jauregui and Jeffrey, Kyle and Jesmonth, Sally and Joshi, Nikhil and Julian, Ryan and Kalashnikov, Dmitry and Kuang, Yuheng and Lee, Kuang-Huei and Levine, Sergey and Lu, Yao and Luu, Linda and Parada, Carolina and Pastor, Peter and Quiambao, Jerry and Rao, Kanishka and Rettinghouse, Jarek and Reyes, Diego and Sermanet, Pierre and Sievers, Nicolas and Tan, Clayton and Toshev, Alexander and Vanhoucke, Vincent and Xia, Fei and Xiao, Ted and Xu, Peng and Xu, Sichun and Yan, Mengyuan},
  journal = {arXiv preprint arXiv:2204.01691},
  year = {2022},
  url = {https://arxiv.org/abs/2204.01691}
}

@inproceedings{liu2023agentbench,
  title = {{AgentBench}: Evaluating {LLMs} as Agents},
  author = {Liu, Xiao and Yu, Hao and Zhang, Hanchen and Xu, Yifan and Lei, Xuanyu and Lai, Hanyu and Gu, Yu and Ding, Hangliang and Men, Kaiwen and Yang, Kejuan and Zhang, Shudan and Deng, Xiang and Zeng, Aohan and Du, Zhengxiao and Zhang, Chenhui and Shen, Sheng and Zhang, Tianjun and Su, Yu and Sun, Huan and Huang, Minlie and Dong, Yuxiao and Tang, Jie},
  booktitle = {International Conference on Learning Representations},
  year = {2024},
  url = {https://arxiv.org/abs/2308.03688}
}

@inproceedings{qin2023toollm,
  title = {{ToolLLM}: Facilitating Large Language Models to Master 16000+ Real-world {APIs}},
  author = {Qin, Yujia and Liang, Shihao and Ye, Yining and Zhu, Kunlun and Yan, Lan and Lu, Yaxi and Lin, Yankai and Cong, Xin and Tang, Xiangru and Qian, Bill and Zhao, Sihan and Tian, Runchu and Xie, Ruobing and Zhou, Jie and Gerstein, Mark and Li, Dahai and Liu, Zhiyuan and Sun, Maosong},
  booktitle = {International Conference on Learning Representations},
  year = {2024},
  url = {https://arxiv.org/abs/2307.16789}
}

@inproceedings{park2023generativeagents,
author = {Park, Joon Sung and O'Brien, Joseph C. and Cai, Carrie J. and Morris, Meredith Ringel and Liang, Percy and Bernstein, Michael S.},
title = {Generative Agents: Interactive Simulacra of Human Behavior},
year = {2023},
publisher = {Association for Computing Machinery},
address = {New York, NY, USA},
booktitle = {In the 36th Annual ACM Symposium on User Interface Software and Technology (UIST '23)},
location = {San Francisco, CA, USA},
series = {UIST '23}
}
